\documentclass[11pt,a4paper]{article}

\usepackage{natbib}
\usepackage{amsmath,amsthm,amssymb, thm-restate}
\usepackage{todonotes}
\usepackage{physics}
\usepackage{float}
\usepackage[margin=1in]{geometry}
\usepackage[boxed,vlined,linesnumbered,ruled,resetcount]{algorithm2e}
\usepackage{stmaryrd}
\usepackage{systeme}
\usepackage{authblk}
\usepackage{tocbibind}
\usepackage[toc,page]{appendix}
\usepackage{bbold}
\usepackage{bm}
\usepackage{comment}
\usepackage{rotating}
\usepackage{pdflscape}
\usepackage{physics}
\usepackage{mathtools}
\usepackage{soul}

\usepackage{hyperref}
\hypersetup{
colorlinks=true,
breaklinks=true,
urlcolor= blue,
linkcolor= black, 
citecolor=black}

\newcommand{\R}{\mathbb{R}}
\newcommand{\Z}{\mathbb{Z}}

\newcommand{\Pcal}{\mathcal{P}}

\newcommand{\OPT}{\textup{OPT}}
\newcommand{\SAT}{D}

\newcommand{\bvect}{\vec{\beta}}

\newcommand{\evect}{\vec{\epsilon}}

\newcommand{\tvect}{\vec{t}}
\newcommand{\sumparam}{\sum p_{ij}}
\newcommand{\Cmax}{C_\textup{max}}

\newcommand{\tshift}{\tau_\text{shift}}

\newcommand{\bigOstar}{\mathcal{O}^*}
\newcommand{\bigO}{\mathcal{O}}

\DeclareMathOperator*{\argmin}{arg\,min}

\newtheorem{theorem}{Theorem}
\numberwithin{theorem}{section}
\newtheorem{definition}[theorem]{Definition}
\newtheorem{remark}[theorem]{Remark}
\newtheorem{example}{Example}

\newtheorem{lemma}[theorem]{Lemma}
\newtheorem{property}[theorem]{Property}

\newtheorem{observation}[theorem]{Observation}

\title{Moderate Exponential-time Quantum Dynamic Programming Across the Subsets for Scheduling Problems\footnote{This is an extension of the conference paper of~\cite{GrangeGecco}.}}
\date{}
\author[1,2]{\underline{Camille Grange}}
\author[1]{Michael Poss}
\author[1]{Eric Bourreau}
\author[3]{Vincent T'kindt}
\author[3]{Olivier Ploton}

\affil[1]{\small{University of Montpellier, LIRMM, CNRS, 161 rue Ada, Montpellier, France}}
\affil[2]{\small{SNCF, Technology, Innovation and Group Projects Department, 1 avenue François Mitterand, Saint-Denis, France}}
\affil[3]{\small{University of Tours, LIFAT, 64 avenue Jean Portalis, Tours, France}}

\begin{document}

\emergencystretch 3em 

\setlength{\parskip}{0mm}
\setlength{\abovedisplayskip}{3mm}
\setlength{\belowdisplayskip}{3mm}
\setlength{\parindent}{4mm}
\allowdisplaybreaks

\maketitle
\setcounter{page}{1}
\renewcommand{\thepage}{\arabic{page}}

\begin{abstract}
    Grover Search is currently one of the main quantum algorithms leading to hybrid quantum-classical methods that reduce the worst-case time complexity for some combinatorial optimization problems. Specifically, the combination of Quantum Minimum Finding (obtained from Grover Search) with dynamic programming has proved particularly efficient in improving the complexity of NP-hard problems currently solved by classical dynamic programming. For these problems, the classical dynamic programming complexity in $\mathcal{O}^*(c^n)$, where $\mathcal{O}^*$ denotes that polynomial factors are ignored, can be reduced by a hybrid algorithm to $\mathcal{O}^*(c_{quant}^n)$, with $c_{quant} < c$. In this paper, we provide a bounded-error hybrid algorithm that achieves such an improvement for a broad class of NP-hard single-machine scheduling problems for which we give a generic description.  
    Moreover, we extend this algorithm to tackle the 3-machine flowshop problem. Our algorithm reduces the exponential-part complexity compared to the best-known classical algorithm, sometimes at the cost of an additional pseudo-polynomial factor.\\
    
    \noindent\textbf{keywords:} Discrete Optimization, Quantum computing, Scheduling, Dynamic Programming
\end{abstract}

\section{Introduction}
The fields of quantum computing and combinatorial optimization are becoming every day more closely linked, thanks to the work of the quantum computing community, as well as the more recent interest of the operations research community that has been focusing on the new quantum paradigm. More precisely, there are two types of quantum algorithms for solving optimization problems. The first type encompasses heuristics, often designed today as hybrid quantum-classical algorithms, such as the class of Variational Quantum Algorithms described by~\cite{Cerezo} or by~\cite{Grange} and, within it, the famous Quantum Approximate Optimization Algorithm (QAOA) of~\cite{Farhi}. Essentially, these algorithms require the optimization problem to be formulated as an unconstrained binary problem with polynomial objective function and can be implemented on current noisy quantum computers because the quantum part can be made rather small. Note that, usually, the problem is formulated as a QUBO (Quadratic Unconstrained Binary Optimization) to limit the number of entanglement gates. Among others, the problems of MAX-CUT~\citep{Farhi}, Travelling Salesman Problem~\citep{TSP}, MAX-3-SAT~\citep{Nannicini_MAX3SAT}, Graph Coloring~\citep{GraphColoring} and Job Shop Scheduling~\citep{JobShop} are reformulated as QUBO and solved with hybrid heuristics on small instances. However, due to the small size of instances processed today and the nature of heuristics whose performances are evaluated empirically, no quantum advantage with these heuristics is emerging yet.
This is where the second type of quantum algorithms differ: they are \emph{exact} algorithms (i.e. that output the optimal solution with a high probability of success) that provide theoretical speed-ups for several types of problems and algorithms, as displayed by~\cite{SimplexQuantum} and~\cite{ConvexDPQuantum}. Notice that with the current quantum hardware, it is impossible to implement them today because of the huge size of quantum resources they require. 

~\cite{Grover} provides one key exact quantum algorithm, that achieves a quadratic speed-up when searching for a specific element in an unsorted table, where the complexity is computed as the number of queries of the table and is done by an oracle. Grover Search represents the routine of many exact quantum algorithms. For instance,~\cite{DurrHoyer} use Grover Search as a subroutine for a hybrid quantum-classical algorithm that finds the minimum of an unsorted table, resulting in the algorithm called Quantum Minimum Finding. Later,~\cite{Ambainis} combine Quantum Minimum Finding with dynamic programming to address NP-hard vertex ordering problems, such as the Traveling Salesman Problem (TSP) or the Minimum Set Cover problem. The problems of interest must satisfy a specific property which implies that they can be solved by classical dynamic programming in $\mathcal{O}^*(c^n)$, where $c$ is usually not smaller than 2. Henceforth, we use $\mathcal{O}^*$ which is the usual asymptotic notation that ignores the polynomial factors in the complexity. The hybrid algorithm of~\cite{Ambainis} reduces the complexity to $\mathcal{O}^*(c_\text{quant}^n)$ for $c_\text{quant} < c$. As an example,~\cite{HeldKarp} dynamic programming solves the TSP in $\mathcal{O}^*(2^n)$ whereas the hybrid algorithm of~\cite{Ambainis} achieves to solve it in $\mathcal{O}^*(1.728^n)$ by combining the dynamic programming recurrence of Held and Karp with Quantum Minimum Finding. Following the work of~\cite{Ambainis}, other NP-hard problems have been tackled with the idea of combining Grover Search (or Quantum Minimum Finding) and classical dynamic programming. This has led to quantum speed-ups for the Steiner Tree problem~\citep{SteinerTree} and the graph coloring problem~\citep{GraphColoring_ProgDyn}. 

The purpose of this work is to provide a hybrid quantum-classical algorithm, adapting the seminal idea of~\cite{Ambainis}, that reduces the time complexity of solving NP-hard scheduling problems. For that, we propose an extended version of well-known Dynamic Programming Across the Subsets (DPAS) recurrences used to solve combinatorial optimization problems like scheduling problems (see e.g.~\cite{TKindt}). Notice that DPAS is a common technique for designing exact algorithms for NP-hard problems as described by~\cite{Woeginger03}. 

\paragraph{Scheduling problems and DPAS.}
\label{sec:intro_scheduling}
A scheduling problem lies in finding the optimal assignment of a set of jobs to machines over time. Each job $j$ is defined by at least a processing time $p_j$ and possibly additional data like a due date $d_j$, a deadline $\Tilde{d}_j$, or even a weight $w_j$ reflecting its priority. One or more machines can process the set of jobs, however, at any time point, a machine can only process one job at a time. The computation of a schedule is done to minimize a given objective function. 

In Sections~\ref{sec:add_dpas} and~\ref{sec:comp_dpas}, we consider single-machine scheduling problems. Let $[n] := \{1, \ldots,n\}$ be the set of jobs to schedule on the machine. While a solution to a single-machine scheduling problem is described by a starting time for each job on the machine, it is standard to describe instead such a solution by a permutation $\pi \in S_{[n]}$ of the $n$ jobs. Indeed, the starting times can be directly deduced from the order of jobs in the permutation and the potential constraints thanks to the following assumptions. First, we assume that only one job can be processed at any time on the machine. Second, we deal only with non-preemptive scheduling, meaning that a job must be run to completion when it started. Henceforth, we use the permutation representation for the solutions. 
In Section~\ref{sec:dec_dpas}, we consider the 3-machine flowshop of $n$ jobs. The definition of this problem, introduced in the above-mentioned section, makes also a solution entirely described by a permutation of $[n]$ even if there are 3 machines. 

Throughout this paper, we use the usual notation $\alpha|\beta|\gamma$, introduced by~\cite{Graham}, to describe the scheduling problem consisting of $\alpha$ machines, with the constraints $\beta$ and the criterion $\gamma$ to be minimized. For instance, $1|\Tilde{d}_j|\sum_j w_jC_j$ is the problem of minimizing the total weighted completion time with deadline constraints on a single machine. 
The reader interested in scheduling can refer to any textbook in scheduling, e.g. to~\cite{Pinedo}.

The single-machine scheduling problems addressed in this work are those that satisfy the Dynamic Programming Across the Subsets (DPAS) property. It means that these problems can be solved by Dynamic Programming where the optimal solution for a set of jobs $J\subseteq [n]$ is computed as the best concatenation over all $j\in J$ of the optimal solution for $J\setminus \{j\}$ and the cost of setting $j$ as the last processed jobs. Specifically, if we note $\OPT[J]$ the optimal value for processing the set $J$ of jobs, the recursion is 
\begin{equation}
\label{eq:DPAS_classical}
    \OPT[J] = \min_{j\in J}\, \OPT[J\setminus \{j\}] + \phi_j\left(\sum_{k\in J} p_k\right)\,,
\end{equation}
where $\phi_j$ is a function depending on job $j$. This generic recursion captures many single-machine scheduling problems as recalled in the survey of~\cite{TKindt}, leading to the worst-case time complexity of $\mathcal{O}^*(2^n)$ to solve all these problems. This naturally raises the question of the existence of moderate exponential-time algorithms with a complexity $\mathcal{O}^*(c^n)$ where $c<2$. The question has been answered positively for specific problems such as minimizing the total weighted completion time with precedence constraints in $\mathcal{O}^*((2-\epsilon)^n)$ for small $\epsilon > 0$ by~\cite{Cygan}. But, as far as we know, no generic method provides such an improvement for a broad class of scheduling problems. In this paper, we present a hybrid algorithm that solves the problems satisfying~\eqref{eq:DPAS_classical} in $\mathcal{O}^*(1.728^n)$, sometimes with an additional pseudo-polynomial factor in the complexity that comes from the generalization of the recurrence. 

\paragraph{Our contributions.} We extend existing recurrences for scheduling problems leading to a quantum speed-up for solving a general class of scheduling problems. The dynamic programming recurrences are adapted to solve scheduling problems with a proposed hybrid algorithm Q-DDPAS, which is an extension of the algorithm of~\cite{Ambainis}. 
Specifically, Q-DDPAS relies on dynamic programming over values indexed by a set, as by~\cite{Ambainis}, but also indexed by a new integer parameter. This new indexation is of a different nature from indexing with a set because it is not exhaustively enumerate in the recurrence but enables to express temporal constraints. Q-DDPAS also deals with non-linear objective functions, for which this is not sufficient anymore to add values of subsets but requires new ways to combine values, e.g. composition, coming from the specificity of scheduling objective functions. Specifically, we cover three types of problems that satisfy three different kinds of dynamic programming properties. For each of them, the best-known classical time complexity is in $\mathcal{O}^*(2^n)$ that is reduced in $\mathcal{O}^*(pseudop\cdot1.728^n)$ by the hybrid algorithm of this paper, where $pseudop$ is a pseudo-polynomial factor. Not only it applies to problems for which the dynamic programming property is based on the \emph{addition} of optimal values of the problem on sub-instances (as done by~\cite{GrangeGecco}) but it also relates to problems for which the dynamic programming naturally applies on the \emph{composition} of optimal values of the problem on sub-instances. Furthermore, we address the 3-machine flowshop problem that differs from previous problems by the nature of the recurrence property and widens the range of problems solved by the hybrid algorithm. Last, we also propose an approximation scheme for the 3-machine flowshop problem based on the hybrid algorithm. 
We summarize in Table~\ref{tab:complexity} the complexities of several NP-hard scheduling problems through which we illustrate the recurrences in this paper.

\begin{table}[ht]
\small
\begin{center}
\begin{tabular}{||c c c||}
 \hline
 Problem & Our hybrid algorithm & Best classical algorithm\\ [1ex] 
 \hline\hline
 $1|\Tilde{d}_j|\sum w_jC_j$ & $\mathcal{O}^* \left(\sum p_j\cdot1.728^n\right)$ & $\mathcal{O}^*(2^n)$~\citep{TKindt}\\ [1ex]
 \hline
 $1||\sum w_jT_j$ & $\mathcal{O}^* \left(\sum p_j\cdot1.728^n\right)$ & $\mathcal{O}^*(2^n)$~\citep{TKindt}\\ [1ex]
 \hline
 $1|prec|\sum w_jC_j$ & $\mathcal{O}^* \left(1.728^n\right)$ & $\mathcal{O}^*((2-\epsilon)^n)$, for small $\epsilon$~\citep{Cygan}\\ [1ex]
 \hline
  $1|r_j|\sum w_jU_j$ & $\mathcal{O}^* \left((\sum w_j)^3\cdot \sum p_j \cdot 1.728^n\right)$ & $\mathcal{O}^*(\sum w_j\cdot \sum p_j\cdot2^n)$~\citep{Ploton}\\ [1ex]
 \hline
 $1|r_j|\sum w_jC_j$ & $\mathcal{O}^* \left((\sum w_j)^3\cdot (\sum p_j)^4 \cdot 1.728^n\right)$ & $\mathcal{O}^*(\sum w_j\cdot (\sum p_j)^2\cdot2^n)$~\citep{Ploton}\\ [1ex]
 \hline
  $F3||C_\text{max}$ & $\mathcal{O}^* \left((\sum p_{ij})^4\cdot 1.728^n\right)$ & $\mathcal{O}^*(3^n)$~\citep{Tkindt_3FlowShop,Ploton_flowshop}\\ [1ex]
 \hline
\end{tabular}
\end{center}
\caption{Comparison of worst-case time complexities between our hybrid algorithm and the best-known classical algorithm.}
 \label{tab:complexity}
 \end{table}

\paragraph{Structure of the paper.}
First, we present in Section~\ref{sec:add_dpas} problems for which the dynamic programming property is based on the \emph{addition} of optimal values of the problem on sub-instances (called Additive DPAS). We begin with the example of problem $1|\Tilde{d}_j|\sum_{j}w_jC_j$ and then provide a generic description of the problems at stake. We describe the related hybrid algorithm (called Q-DDPAS) as it is usually done in the algorithmic quantum literature, namely with a high-level description where quantum \emph{boxes} interact with the classical part. We provide a rigorous and detailed description of the circuit-based implementation for interested readers in our companion paper~\citep{Grange_CompanionPaper}. Similarly, we tackle in Section~\ref{sec:comp_dpas} problems for which the dynamic programming property is based on the \emph{addition} of optimal values of the problem on sub-instances (called Composed DPAS), beginning with the example of problem $1|r_j|\sum w_jU_j$. We provide in Section~\ref{sec:application_scheduling} some applications of Q-DDPAS to the scheduling literature. Lastly, in Section~\ref{sec:dec_dpas}, we consider the 3-machine flowshop problem, for which the dynamic programming recurrence applies to its decision version. It results in a slightly different hybrid algorithm. Additionally, we provide an approximation scheme for this problem, based on the hybrid algorithm, that disposes of the pseudo-polynomial factor in the time complexity. 
We recall in Appendix~\ref{appendix:Bounds} useful bounds to derive the complexities of the proposed algorithms.

\section{Additive DPAS}
\label{sec:add_dpas}
In this section, we present problems for which the dynamic programming recursion is based on the \emph{addition} of optimal values of problems for sub-instances. Next, we detail the hybrid algorithm Q-DDPAS to solve these problems.

\subsection{A scheduling example}
\label{subsec:add_dpas_example}
The NP-hard single-machine scheduling problem at stake is the minimization of the total weighted completion time with deadline constraints, often referred to as $1|\Tilde{d}_j|\sum_{j}w_jC_j$ in the scheduling literature.
The input is given, for each job $j\in[n]$, by a weight $w_j$, a processing time $p_j$ and a deadline $\Tilde{d}_j$ before which the job must be completed. We define the completion time $C_j(\pi)$ of job $j$ as the end time of the job on the machine for the permutation $\pi$. So, if $j$ starts as time $t$ for the permutation $\pi$, then $C_j(\pi) = t + p_j$. 
The problem aims at finding the feasible permutation for which the total weighted sum of completion times is minimal. A permutation $\pi$ is feasible if $C_j(\pi) \leq \Tilde{d}_j$ for all job $j$. 
Thus, the problem can be formulated as follows: 
\begin{equation*}
    \min_{\pi\in\Pi}\, \sum_{j=1}^n w_jC_j(\pi)\,,
\end{equation*}
where the set of feasible permutations is $
    \Pi = \{\pi \in S_{[n]}\,|\, C_j(\pi) \leq \Tilde{d}_j\,, \forall j\in [n]\}\,.$

This problem satisfies two recurrences. For deriving them, we need to introduce the set $T := \left\llbracket0,\sum_{j=1}^n p_j\right\rrbracket$, where we use the notation $\llbracket a,b\rrbracket = \{a,a+1,\ldots,b\}$ for integers $a$ and $b$. For $J\subseteq[n]$ and $t\in T$, we define $\OPT[J,t]$ as the optimal value of the problem in which only jobs in $J$ are scheduled from time $t$. Thus, solving our nominal problem $1|\Tilde{d}_j|\sum_{j}w_jC_j$ amounts to compute $\OPT[[n],0]$.

The first recurrence comes from the standard Dynamic Programming Across the Subsets (DPAS) described in~\eqref{eq:DPAS_classical}. However, compared to usual DPAS, we introduce an extra parameter $t$ necessary for the solution with our hybrid algorithm as explained later. The idea of this recurrence is to get the optimal value of our problem for jobs in $J$ and starting at time $t$ by finding, over all jobs $j\in J$, the permutation that ends by $j$ with the best cost value. It is possible to do so because no matter what the optimal permutation of the first $(|J|-1)$ jobs is, the cost of setting job $j$ at the end of the permutation is always known. Indeed, the time taken to process all jobs in $J\setminus\{j\}$ is always $\sum_{k\in J\setminus\{j\}}p_k$. Thus, the completion time of $j$ is defined by $c_j = t+\sum_{k\in J}p_k$. It results that the cost of setting $j$ at the end of the permutation is $w_j(t+\sum_{k\in J}p_k)$. It also implies that the deadline constraint for job $j$ is satisfied if $t + \sum_{k\in J}p_k \leq \Tilde{d}_j$. Specifically, for all $J \subseteq [n]$ and for all $t\in T$, we have 
\begin{equation}
\label{eq:add_dpas_ex}
    \OPT[J,t] =
    \min\limits_{j\in J}\, \OPT[J\setminus \{j\},t] + \left\{
     \begin{aligned}
      & w_j\left(t + \sum_{k\in J}p_k\right)~~ &\text{if } t + \sum_{k\in J}p_k \leq \Tilde{d}_j\\
      & +\infty~~ &\text{otherwise}
    \end{aligned}
  \right.\,,
\end{equation}
initialized by $\OPT[\emptyset] = 0$.

The second recurrence generalizes the previous one. 
For this recurrence, the principle of computing $\OPT[J,t]$ is similar to~\eqref{eq:add_dpas_ex} but instead of setting one job at the end of the permutation, we choose $|J|/2$ jobs and set them to be the half last jobs of the permutation. Specifically, for all $J\subseteq[n]$ of even cardinality and $t\in T$, we have
\begin{equation}
\label{eq:add_d_dpas_ex}
\OPT[J,t] = \min\limits_{X\subseteq J\atop
|X|=|J|/2} \Bigl\{\OPT[X, t] + \OPT[J\setminus X, t + \sum_{i\in X}p_i]\Bigr\}\,,
\end{equation}
initialized by, $\forall j \in [n]$ and $t\in T$, $\OPT[\{j\},t] = \left\{\begin{aligned}
      & w_j(p_j + t)~~ &\text{if } \Tilde{d}_j \geq p_j + t\\
      & +\infty~~ &\text{otherwise}
    \end{aligned}\right.\,.$
    
For a given $X\subseteq J$ of size $|J|/2$, recurrence~\eqref{eq:add_d_dpas_ex} computes the best permutation of jobs in $X$ starting at time $t$, and the best permutation of jobs in $J\setminus X$ starting at time $t+\sum_{k\in X}p_k$ as we know that, as before, no matter what is the optimal permutation for jobs in $X$, the time taken to process them all is exactly $\sum_{k\in X}p_k$.

The two above recurrences have been illustrated with problem $1|\Tilde{d}_j|\sum_{j}w_jC_j$. In the next subsection, we propose a general formulation of these recurrences that will be used to elaborate our algorithm as general as possible to solve a broad class of scheduling problems.

\subsection{General formulation of recurrences}
\label{subsec:add_dpas_general}
Let us consider the following general scheduling problem:
\begin{equation*}
    \Pcal : \hspace{6mm} \min_{\pi\in \Pi} f(\pi)\,,
\end{equation*}
where $\Pi \subseteq S_{[n]}$ is the set of feasible permutations of $[n] := \{1,\ldots,n\}$ according to given constraints and $f$ is the objective function. We introduce a related problem $P$ useful for deriving the dynamic programming recursion, for which we specify the instance: for $J\subseteq [n]$ and $t\in \Z$,
\begin{equation}
\label{eq:related_problem}
    P(J,t) : \hspace{6mm} \min_{\pi \in \Pi(J,t)} f(\pi,J,t)
\end{equation} as the nominal scheduling problem $\Pcal$ that schedules only jobs in $J$ and starts the schedule at time $t$. Let us note $\OPT[J,t]$ the optimal value of $P(J,t)$.
It results that solving $\Pcal$ amounts to solving $P([n],0)$, and it can be performed by Q-DDPAS if the related problem $P$ satisfies the two recurrences~\eqref{eq:add_dpas_recu} and~\eqref{eq:add_d_dpas_recu} below. Henceforth, we denote by $2^{[n]}$ the set of all subsets of $[n]$, and by $\llbracket a,b\rrbracket$ the set $\{a, a+1,\ldots,b\}$. Let us introduce the first recurrence.

\begin{property}[Additive DPAS]
\label{prop:Add-DPAS}
There exists a function $g : 2^{[n]}\times [n]\times T \rightarrow \mathbb{R}$, computable in polynomial time, such that, for all $J\subseteq [n]$ and for all $t_0\in T$,
\begin{equation}
\tag{Add-DPAS}
\label{eq:add_dpas_recu}
    \OPT[J,t_0] =
    \min\limits_{j\in J}  \Bigl\{\OPT[J\setminus \{j\}, t_0] + g(J,j,t_0) \Bigr\}
\end{equation}
initialized by $\OPT[\emptyset, t_0] = 0$.
\end{property}
\begin{lemma}
\label{lemma:Add_DPAS_complexity}
    Dynamic programming~\eqref{eq:add_dpas_recu} solves $\Pcal$ in  $\bigOstar(2^n)$.
\end{lemma}
\proof
We solve Equation~\eqref{eq:add_dpas_recu} for all $J$ such that $|J| = k$, and for $t_0 = 0$, starting from $k=1$ to $k=n$. 
For a given $J$, the values $\{\OPT[J\setminus \{j\},0] : j\in J\}$ are known, so $\OPT[J,0]$ is computed in time $poly(n)\cdot k$ according to Equation~\eqref{eq:add_dpas_recu}, where $poly(n)$ is a polymial function of $n$ (the computation of $g$ is polynomial).
Eventually, the total complexity of computing $\OPT[[n],0]$ is $\sum_{k=1}^n poly(n)k\binom{n}{k} =  poly(n)\cdot n\cdot2^{n-1} =  \mathcal{O}^*(2^n).$
\endproof

Throughout, we commit a slight abuse of language by letting~\eqref{eq:add_dpas_recu} both refer to the property satisfied by a given optimization problem and to the resulting dynamic programming algorithm. Notice the presence of the additional parameter $t_0$ in the above definition, which is typically absent in the scheduling literature. In particular, $t_0$ is a constant throughout the whole recursion~\eqref{eq:add_dpas_recu} and does not impact the resulting computational complexity. The use of that extra parameter in $T$ shall be necessary later when applying our hybrid algorithm. 

Property~\ref{prop:Add-DPAS} expresses that finding the optimal value of $P$ for jobs in $J$ and starting at time $t$ is done by finding over all jobs $j\in J$ the permutation that ends by $j$ with the best cost value. Function $g$ represents the cost of $j$ being the last job of the permutation. Notice that isolating the last job of the permutation is a usual technique in scheduling as displayed in~\eqref{eq:DPAS_classical}. In the second recurrence below, we provide a similar scheme, where instead of one job, we \emph{isolate} half of the jobs in $J$, turning the computation of $g$ to the solution of another problem on a sub-instance with $|J|/2$ jobs. 

\begin{property}[Additive Dichotomic DPAS]
\label{prop:Add-D-DPAS}
There exist two functions $\tshift : 2^{[n]}\times 2^{[n]}\times T \rightarrow T$ and $h : 2^{[n]}\times 2^{[n]}\times T \rightarrow \mathbb{R}$, computable in polynomial time, such that, for all $J\subseteq [n]$ of even cardinality, and for all $t\in T$,
\begin{align}
\tag{Add-D-DPAS}
\label{eq:add_d_dpas_recu}
\OPT[J,t] = \min\limits_{X\subseteq J\atop
|X|=|J|/2} \Bigl\{\OPT[X, t] + h(J,X,t) + \OPT[J\setminus X,\tshift(J,X,t)]\Bigr\}
\end{align}
initialized by the values $\OPT[\{j\}, t]$ for each $j\in [n]$ and $t\in T$.
\end{property}
For a given $X\subseteq J$, the above recursion computes the best permutation of jobs in $X$ starting at time $t$, and the best permutation of jobs in $J\setminus X$ starting at time $\tshift$, adding the function $h$ that represents the cost of the concatenation between these two permutations.

\begin{remark}
\label{rem:recu_equiv}
We observe that problem~\eqref{eq:related_problem} satisfies recurrence~\eqref{eq:add_dpas_recu} if and only if it satisfies~\eqref{eq:add_d_dpas_recu}. This can be seen by developing recursively both recurrences, which essentially leads to optimization problems over $\pi\in S_{[n]}$, whose objective functions respectively involve $g$ in the first case and $h$ and $\tshift$ in the second case. Here, one readily verifies that $g$ can then be defined from $h$ and $\tshift$ and reciprocally.
\end{remark}

Despite the previous remark, the two recurrences differ on the size of the subsets considered along the recursions, leading to different formulations and therefore require more or less sub-problems to be solved optimally in the dynamic programming process. This is formalized in the following proposition. Note that we use the notation $f_1(n) = \omega(f_2(n))$ if $f_1$ dominates asymptotically $f_2$.
\begin{restatable}{lemma}{AddDDPAScomplexity}
\label{lem:Add_D_DPAS_complexity}
    Dynamic programming~\eqref{eq:add_d_dpas_recu} solves $\Pcal$ in  $\omega(|T|\cdot 2^n)$.
\end{restatable}
The proof is given in the Supplementary Materials.
Notice that if $n$ is not a power of 2, we can still add fake jobs without changing the following conclusion: solving $\Pcal$ with~\eqref{eq:add_dpas_recu} is faster than with~\eqref{eq:add_d_dpas_recu}. However, in the next subsection, we describe a hybrid algorithm Q-DDPAS that improves the complexity of solving $\Pcal$ by combining recurrences~\eqref{eq:add_dpas_recu} and~\eqref{eq:add_d_dpas_recu} with a quantum subroutine. 

\subsection{Hybrid algorithm for Additive DPAS}
In this subsection, we describe our hybrid algorithm Q-DDPAS adapted from the work of~\cite{Ambainis}. Notice that it assumes to have a quantum random access memory (QRAM)~\citep{QRAM}, namely, to have a classical data structure that stores classical information but can answer queries in quantum superposition. We underline that this latter assumption is strong because QRAM is not yet available on current universal quantum hardware. 
First, let us introduce the Quantum Minimum Finding algorithm of~\cite{DurrHoyer}, which constitutes a fundamental subroutine in our algorithm. This algorithm essentially applies several times Grover Search~\citep{Grover} and provides a quadratic speedup for the search of a minimum element in an unsorted table. 

\begin{definition}[Quantum Minimum Finding~\citep{DurrHoyer}]
\label{def:QMF}
Let $f:[n]\rightarrow \Z$ be a function. 
Quantum Minimum Finding computes the minimum value of $f$ and the corresponding minimizer $\argmin_{i\in [n]} \{f(i)\}$.
The complexity of Quantum Minimum Finding is $\mathcal{O}\left(\sqrt{n}\cdot C_f(n)\right)$, where $\mathcal{O}(C_f(n))$ is the complexity of computing a value of $f$.
\end{definition}

\begin{remark}[Success probability and bounded-error algorithm~\citep{Bernstein93}]
\label{rem:bounded_error}
~\cite{DurrHoyer} prove that Quantum Minimum Finding computes the minimum value with a probability of success strictly larger than $\frac{1}{2}$, independent of $n$. Thus, for $\epsilon>0$, finding the minimum value with probability $(1-\epsilon)$ is achieved by repeating $\mathcal{O}(\log \frac{1}{\epsilon})$ times Quantum Minimum Finding. Henceforth, we refer to this statement when we write that Quantum Minimum Finding finds the minimum value \emph{with high probability}. Equivalently, we say that this is a \emph{bounded-error} algorithm. 
More generally, in the rest of the paper, we call a \emph{bounded-error algorithm} an algorithm that provides the optimal solution with a probability as close to 1 as we want by repeating it a number of times independent of the instance size.
\end{remark}

Next, we describe the algorithm of~\cite{Ambainis} adapted for our Additive DPAS recurrences which implies extra parameters in $T$. We call it Q-DDPAS and it consists essentially of calling recursively twice Quantum Minimum Finding and computing classically the left terms. Without loss of generality, we assume that 4 divides $n$. This can be achieved by adding at most three fake jobs and, therefore, does not change the algorithm complexity. Q-DDPAS consists of two steps. First, we compute classically by~\eqref{eq:add_dpas_recu} the optimal values of $P$ on sub-instances of $n/4$ jobs and for all starting times $t\in T$. Second, we call recursively two times Quantum Minimum Finding with~\eqref{eq:add_d_dpas_recu} to find optimal values of $P$ on sub-instances of $n/2$ jobs starting at any time $t\in T$, and eventually of $n$ jobs starting at $t=0$ (corresponding to the optimal value of the nominal problem $\Pcal$). Specifically, we describe Q-DDPAS in Algorithm~\ref{alg:QDDPAS}.
\begin{algorithm}[!ht]
\label{alg:QDDPAS}
\LinesNotNumbered
\KwIn{Problem $P$ satisfying~\eqref{eq:add_dpas_recu} and~\eqref{eq:add_d_dpas_recu}}
\KwOut{$\OPT[[n],0]$ with high probability}
\caption{Q-DDPAS for Additive DPAS}
\Begin(\textbf{classical part}){
\For{$X \subseteq [n]$ such that $|X| = n/4$, and $t\in T$}{
\nl \label{ligne:classic} Compute $\OPT[X,t]$ with~\eqref{eq:add_dpas_recu} and store the results in the QRAM\;
}}
\Begin(\textbf{quantum part}){
\nl \label{ligne:quant1} Apply Quantum Minimum Finding with~\eqref{eq:add_d_dpas_recu} to find $\OPT[[n],0]$\;
\nl \label{ligne:quant2} To get values for the Quantum Minimum Finding above (the values $\OPT[J,t]$ for $J\subseteq[n]$ of size $n/2$ and $t\in T$), apply Quantum Minimum Finding with~\eqref{eq:add_d_dpas_recu}\;
\nl \label{ligne:quant3} To get values for the Quantum Minimum Finding above (the values $\OPT[X,t']$ for $X\subseteq[n]$ of size $n/4$ and $t'\in T$), get them on the QRAM
}
\end{algorithm}

\begin{theorem}
\label{thm:QDDPAS}
    The bounded-error algorithm Q-DDPAS (Algorithm~\ref{alg:QDDPAS}) solves $\mathcal{P}$ in $\mathcal{O}^*(|T|\cdot1.754^n)$.
\end{theorem}

The detailed proof of the correctness of the algorithm, involving the description of the gate implementation, is detailed in the companion paper~\cite{Grange_CompanionPaper}, with all the low-level details for implementing the algorithm. 

\proof
Hereafter, we provide a high-level proof. The upper bounds to simplify the complexities are detailed in the Appendix~\ref{appendix:Bounds}.
\begin{itemize}
    \item Classical part: computing all $\OPT[X,t]$ for all $X$ of size $n/4$ and for all $t\in T$ (Step~\ref{ligne:classic} in Algorithm~\ref{alg:QDDPAS}) is done by~\eqref{eq:add_d_dpas_recu} in time $\mathcal{O}^*\left(|T|\cdot\sum_{k=1}^{n/4}k\binom{n}{k}\right) = \mathcal{O}^*(|T|\cdot 2^{0.811n})$.
    \item Quantum part: according to Quantum Minimum Finding complexity (Definition~\ref{def:QMF}), computing $\OPT[[n],0]$ with Quantum Minimum Finding (Step~\ref{ligne:quant1} in Algorithm~\ref{alg:QDDPAS}) is done in $\bigO\left(\sqrt{\binom{n}{n/2}}\cdot C_1(n)\right)$, where $C_1(n)$ is the complexity of computing $\OPT[J,t]$ for a $J$ of size $n/2$ and $t\in T$. \emph{The essence of the quantum advantage here is that we do not need to enumerate all sets $J$ and all time $t$ but we apply the Quantum Minimum Finding in parallel to all at once.} Notice that $\binom{n}{n/2}$ is the number of balanced bi-partitions of $[n]$, namely the number of elements we search over to find the minimum of Equation~\eqref{eq:add_d_dpas_recu} when computing $\OPT[[n],0]$. Thus, $C_1(n)$ is exactly the complexity of Quantum Minimum Finding applied on Step~\ref{ligne:quant2} in Algorithm~\ref{alg:QDDPAS}, namely $C_1(n) = \bigO\left(\sqrt{\binom{n/2}{n/4}}\cdot C_2(n)\right)$ where $C_2(n)$ is the complexity of computing $\OPT[X,t']$ for $X$ of size $n/4$ and $t'\in T$. Those values are already computed and stored in the QRAM (Step~\ref{ligne:classic} in Algorithm~\ref{alg:QDDPAS}), namely, $C_2(n) = \bigOstar(1)$.
Thus, the quantum part complexity is $\mathcal{O}^*\left(\sqrt{\binom{n}{n/2}\binom{n/2}{n/4}}\right) = \mathcal{O}^*(2^{0.75n})$.
\end{itemize} 
Eventually, Q-DDPAS complexity is the maximum of the classical and the quantum part complexity. Specifically, the total complexity is $\mathcal{O}^*(|T|\cdot 2^{0.811n}) = \mathcal{O}^*(|T|\cdot 1.754^n)$.
\endproof

We observe that the complexity of Q-DDPAS can be further reduced by performing a third call to Equation~\eqref{eq:add_d_dpas_recu} as suggested by~\cite{Ambainis}.
\begin{restatable}{observation}{QDDPASimprovement}
\label{obs:QDDPAS_modif}
    A slight modification of Q-DDPAS reduces the complexity to $\mathcal{O}^*(|T|\cdot 1.728^n)$.
\end{restatable}
\proof
The slight modification of Q-DDPAS amounts to adding a level of recurrence in the quantum part so that the complexity of the classical part reduces whereas the complexity of the quantum part increases so that both are equal and thus minimize the total complexity. The third call searches for the best concatenation among all the bi-partitions of size $(0.945\cdot \frac{n}{4}, 0.055\cdot \frac{n}{4})$ (that are integers asymptotically), i.e. solving
\begin{align*}
\OPT[J,t] = \min\limits_{X\subseteq J\atop
|X|=0.945|J|} \Bigl\{\OPT[X, t] + h(J,X,t) + \OPT[J\setminus X,\tshift(J,X,t)]\Bigr\}\,.
\end{align*}
The classical part computes all $\OPT[X,t]$ for $X$ of size $0.945\cdot\frac{n}{4}$ and $0.055\cdot\frac{n}{4}$, in $\bigOstar(1.728^n)\,.$ The quantum part applies three levels of recurrence of Quantum Minimum Finding, computing the minimum over functions with a domain of size $\binom{n}{n/2}$, $\binom{n/2}{n/4}$ and $\binom{n/4}{0.945\cdot n/4}$ respectively. Its complexity is then $\bigOstar\left(\sqrt{\binom{n}{n/2}\binom{n/2}{n/4}\binom{n/4}{0.945\cdot n/4}}\right) = \bigOstar(1.728^n)$ (see Appendix~\ref{appendix:Bounds}). 
\endproof

Notice that the classical part of Q-DDPAS can be replaced by any classical algorithm $\mathcal{A}$, if $\mathcal{A}$ computes in $\mathcal{O}^*(|T|\cdot 1.728^n)$ all $\OPT[X,t]$ for $X\subseteq[n]$ of size $n/4$ and $t\in T$. Moreover, if $\mathcal{A}$ happens to reduce the classical part complexity $\mathcal{O}^*(|T|\cdot c^n)$ for $c<1.728$, the complexity of Q-DDPAS can also be reduced in the same spirit as the slight modification of Observation~\ref{obs:QDDPAS_modif}.

The application of Q-DDPAS for Additive DPAS to the specific problem  $1|\Tilde{d}_j|\sum_{j}w_jC_j$ introduced in Subsection~\ref{subsec:add_dpas_example} is given in Subsection~\ref{subsec:scheduling_add_dpas}, together with other scheduling examples. Before introducing other types of problems tackled by Q-DDPAS in the next section, we provide some insights to underline why the use of the quantum subroutine Quantum Minimum Finding in Q-DDPAS must be carefully combined with classical computation to achieve a quantum speedup.

\begin{remark}
    Solving $\Pcal$ with~\eqref{eq:add_dpas_recu} and replacing each classical computation of the minimum by the quantum subroutine Quantum Minimum Finding would not improve the best classical complexity. Indeed, the complexity would be $\sum_{k=1}^n poly(n)\sqrt{k}\binom{n}{k} =  \mathcal{O}^*(2^n)\,.$
\end{remark}

\begin{remark}
    Solving $\Pcal$ exclusively by recursive calls to Quantum Minimum Finding (thus avoiding the classical computations for sets of size $n/4$) would not improve the classical complexity. Using recurrence~\eqref{eq:add_d_dpas_recu}, which is the quantum part of Algorithm~\ref{alg:QDDPAS} with roughly $\log_2(n)$ recursive calls, would give a complexity in $\bigO\left(\sqrt{\binom{n}{n/2}\binom{n/2}{n/4}\ldots\binom{2}{1}}\right)$ that is worse than $\bigOstar(2^n)$. Using recurrence~\eqref{eq:add_dpas_recu} would be even worse because it would require $n$ recursive calls leading to the complexity $\bigO(\sqrt{n(n-1)\ldots 1})$.
\end{remark}


\section{Composed DPAS}
\label{sec:comp_dpas}

In this section, we study scheduling problems whose constraints enable only the \emph{composition} of problems on sub-instances. We describe the adaptation of Q-DDPAS for these problems.

\subsection{A scheduling example}
\label{subsec:comp_dpas_example}
We begin with the specific problem of minimizing the total weighted number of late jobs with release date constraints, often referred to as $1|r_j|\sum w_jU_j$ in the literature. The input is given by, for each job $j\in [n]$, a weight $w_j$, a processing time $p_j$, a release date $r_j$ that is the time from which the job can be scheduled (and not before), and a due date $d_j$ that indicates the time after which the job is late. Thus, a job $j$ is late in permutation $\pi$ if its completion time is larger than $d_j$, namely if $C_j(\pi) > d_j$. We name $U_j(\pi) = \mathbb{1}_{C_j(\pi) > d_j}$ its indicator function of lateness. This problem aims at finding the feasible permutation, namely where each job starts after its release date, for which the total weighted number of late jobs is minimal. Thus, the problem can be formulated as follows:
\begin{equation*}
    \min_{\pi\in \Pi}\, \sum_{j=1}^n w_jU_j(\pi)\,,
\end{equation*}
where the set of feasible solutions is $\Pi = \{\pi\in S_{[n]}\,|\, C_j(\pi) \geq r_j + p_j\}\,.$

This problem does not satisfy the recurrences~\eqref{eq:add_dpas_recu} and~\eqref{eq:add_d_dpas_recu} because the release date constraints do not allow the addition of sub-instances. Let us take the example of~\eqref{eq:add_d_dpas_recu}. The starting time of the second half of jobs $J\setminus X$ in~\eqref{eq:add_d_dpas_recu} can be known only if we know the optimal permutation of the first half job, which is in opposition with the dynamic programming principle. Indeed, the release dates enable empty slots in the scheduling on the first half job such that the time to process all these jobs is not always equal to $\sum_{k\in X} p_k$ and can be larger. 

This observation leads to different recurrences, where the time to process the jobs would be known by dynamic programming. For that, we define an auxiliary problem on which the recurrences apply and we introduce a new set of parameters $E := \left\llbracket0,\sum_{j=1}^n w_j\right\rrbracket$. For $J\subseteq[n]$, $t\in T := \left\llbracket0,\sum_{j=1}^n p_j\right\rrbracket\cup \{+\infty\}$ and $\epsilon\in E$, we note $\OPT[J,t,\epsilon]$ the minimum makespan, i.e. the completion time of the last job, for jobs in $J$ beginning at time $t$ where the weighted number of late jobs is exactly $\epsilon$. Notice that by convention, $\OPT[J,t,\epsilon] = +\infty$ if there is no feasible solution, i.e. if $\left\{\pi\in S_J : C_j(\pi)\geq \max(t,r_j) + p_j\,, \forall j\in J \mbox{ and } \sum_{j\in J} w_jU_j(\pi) = \epsilon\right\} = \emptyset$. Thus, our initial problem $1|r_j|\sum w_jU_j$ is 
$$\min\limits_{\epsilon\in E} \{\epsilon : \OPT[[n],0,\epsilon] < +\infty\}\,.$$
The following recurrence that satisfies the auxiliary problem is inspired by the work of~\cite{Lawler} for the problem of minimizing the total weighted number of late jobs on a single machine under preemption and release date constraints ($1|r_j, pmtn|\sum w_jU_j$). For $J\subseteq [n]$, $t\in T$, $\epsilon\in E$,
\begin{equation*}
    \OPT[J,t,\epsilon] = \min\limits_{j\in J} \Bigl\{ \underbrace{\OPT\Big[\{j\}, \OPT[J\setminus \{j\},t,\epsilon],0\Big]}_{\text{job $j$ is not late}},  \underbrace{\OPT\Big[\{j\}, \OPT[J\setminus \{j\},t,\epsilon - w_j],w_j\Big]}_{\text{job $j$ is late}}\Bigr\}\,.
\end{equation*}
In this recurrence, for each $j\in J$, we impose $j$ as the last job of the permutation and distinguish two cases, whether it is late or not. Notice that the starting time of $j$ is known and equal to $\OPT[J\setminus \{j\},t,.]$ which represents the value for the time parameter. 
The recurrence is initialized by, for $j\in[n]$, $t\in T$ and $\epsilon\in E$,
\begin{equation*}
\OPT[\{j\},t,\epsilon] =
    \begin{cases}
        C_j := \max(t,r_j)+p_j,~~\text{ if } C_j \leq d_j \text{ and } \epsilon = 0\\
        +\infty ,~~\text{ if } C_j> d_j \text{ and } \epsilon = 0\,, \text{ or if } C_j\leq d_j \text{ and } \epsilon = w_j\\
        C_j ,~~\text{ if } C_j> d_j \text{ and } \epsilon = w_j\\
        +\infty ,~~\text{ if } \epsilon \in \llbracket1,w_j - 1\rrbracket\cup\llbracket w_j + 1,\mathlarger\sum_{k=1}^n w_k\rrbracket
    \end{cases}
\end{equation*}

This recurrence generalizes into the following \emph{dichotomic} version for which, instead of setting the last job of the permutation, we set the half-last jobs. For all $J \subseteq [n]$ of even cardinality, $t\in T$ and $\epsilon\in E$,
\begin{equation*}
    \OPT[J,t,\epsilon] = \min\limits_{\epsilon'\in E\atop X\in J : |X|=|J|/2} \Bigl\{ \OPT\Big[X, \OPT[J\setminus X,t,\epsilon - \epsilon'],\epsilon'\Big]\Bigr\}\,,
\end{equation*}
initialized by the same values of $\OPT[\{j\},t,\epsilon]$ for $j\in[n]$, $t\in T$ and $\epsilon\in E$.
Next, we provide generic recurrences to consider problems for which the composition of sub-instances is possible.

\subsection{General formulation of recurrence}
\label{subsec:comp_dpas_general}
Let us consider a scheduling problem with $n$ jobs $$\Pcal : \hspace{6mm} \min_{\pi\in \Pi} f(\pi)\,,$$ 
where $\Pi \subseteq S_{[n]}$ is the set of feasible permutations of $[n] := \{1,\ldots,n\}$ according to given constraints and $f$ is the objective function. Following the example detailed in the previous subsection, we consider an auxiliary problem $\Pcal'$ useful for deriving the dynamic programming recursion, for which we specify the instance: for $J\subseteq [n]$ the jobs to be scheduled, $t\in \Z$ the starting time of the schedule and $\epsilon\in \Z$, we define 
\begin{equation}
\label{eq:related_auxiliary_problem}
    P'(J,t,\epsilon) : \hspace{6mm} \min_{\pi \in \Pi'(J,t,\epsilon)} f'(\pi,J,t,\epsilon)\,,
\end{equation} 
where $f'$, respectively $\Pi'$, is the objective function, respectively the feasible set, are different from those of $\Pcal$. 
We assume that solving $\Pcal$ amounts to finding the smallest $\epsilon\in\Z$ such that the auxiliary problem $\Pcal'$ is bounded. Specifically, 
\begin{equation}
\label{eq:def_p_comp_cal}
    \Pcal : ~~ \min_{\epsilon\in \Z} \Bigl\{\epsilon : \OPT[[n],0,\epsilon] < +\infty\Bigr\}\,.
\end{equation}

To solve the nominal problem $\mathcal{P}$ by classical dynamic programming, 
problem $\Pcal'$ must satisfy recurrence~\eqref{eq:comp_dpas_recu} or recurrence~\eqref{eq:comp_d_dpas_recu} below (as in Remark~\ref{rem:recu_equiv}, we can state that a problem satisfies one if and only if it satisfies the other one). As we explain later, solving $\mathcal{P}$ with our hybrid algorithm requires problem $P'$ to satisfy the two recurrences.

\begin{property}[Composed DPAS]
For all $J \subseteq [n]$, $t\in T$ and $\epsilon\in E$,
\begin{equation}
\label{eq:comp_dpas_recu}
\tag{Comp-DPAS}
    \OPT[J,t,\epsilon] = \min\limits_{\epsilon'\in E\atop j\in J} \Bigl\{ \OPT\Big[\{j\}, \OPT[J\setminus \{j\},t,\epsilon - \epsilon'], \epsilon'\Big]\Bigr\}\,,
\end{equation}
initialized by the values of $\OPT[\{j\},t,\epsilon]$ for all $j\in[n]$, $\epsilon\in E$ and $t\in T$. Notice that for $J\subseteq [n]$, $t\in T$ and $\epsilon\in E$, we adopt the convention $\OPT[J,t,\epsilon] = +\infty$ for $\epsilon \notin E$.
\end{property}
Recurrence~\eqref{eq:comp_dpas_recu} differs from recurrence~\eqref{eq:add_dpas_recu} in two aspects. 
First, the optimal values of the problem on sub-instances are composed, and not added, because of the nature of the constraints. Second, the search for the minimum value is done not only over all jobs in $J$, but also over all values in $E$. More precisely, for a given $\epsilon_0 \in E$, the optimal value of $P'(J,t,\epsilon_0)$ is the minimum value of all possible composition of optimal values of the problem on sub-instances with parameters $\epsilon_1$ and $\epsilon_2$ such that $\epsilon_1 + \epsilon_2 = \epsilon_0$. We have the following result.
\begin{lemma}
\label{lemma:Comp_DPAS_complexity}
    ~\eqref{eq:comp_dpas_recu} solves $\mathcal{P}$ in $\mathcal{O}^*(|E|^3\cdot |T|\cdot2^n)$. 
\end{lemma}
\proof
Let $\epsilon_0 \in E$. Similarly to the proof of Lemma~\ref{lemma:Add_DPAS_complexity}, we show that~\eqref{eq:comp_dpas_recu} solves $P'([n],0,\epsilon_0)$ in $\mathcal{O}^*(|E|^2\cdot |T| \cdot2^n)$. Indeed, to compute $\OPT[[n],0,\epsilon_0]$, we need to solve 
Equation~\eqref{eq:comp_dpas_recu} for all $J$ such that $|J| = k$ starting from $k=1$ to $k=n$, and for all $t\in T$ and $\epsilon\in E$. For a given $J$, $t\in T$ and $\epsilon\in E$, the values $\{\OPT[J\setminus \{j\},t',\epsilon'] : j\in J, t'\in T, \epsilon'\in E\}$ and  $\{\OPT[ \{j\},t',\epsilon'] : j\in J, t'\in T, \epsilon'\in E\}$ are known, so $\OPT[J,t,\epsilon]$ is computed in time $|E|\cdot k$ according to Equation~\eqref{eq:comp_dpas_recu}. Eventually, the total complexity of computing $\OPT[[n],0,\epsilon_0]$ is 
$\sum_{k=1}^n |T|\cdot |E|^2\cdot k\binom{n}{k} =  \mathcal{O}^*(|T|\cdot |E|^2\cdot2^n).$
Moreover, solving $\mathcal{P}$ amounts to solving $P'([n],0,\epsilon)$ ,for all $\epsilon\in E$, according to~\eqref{eq:def_p_comp_cal}. The complexity results directly from the above complexity of computing $\OPT[[n],0,\epsilon_0]$, for $\epsilon_0 \in E$.
\endproof

The auxiliary problem $P'$ must satisfy the following recurrence~\eqref{eq:comp_d_dpas_recu} in addition to recurrence~\eqref{eq:comp_dpas_recu}. 

\begin{property}[Composed Dichotomic DPAS]
For all $J \subseteq [n]$ of even cardinality, $t\in T$ and $\epsilon\in E$,
\begin{equation}
\tag{Comp-D-DPAS}
\label{eq:comp_d_dpas_recu}
    \OPT[J,t,\epsilon] = \min\limits_{\epsilon'\in E\atop X\in J : |X|=|J|/2} \Bigl\{ \OPT\Big[X, \OPT[J\setminus X,t,\epsilon - \epsilon'],\epsilon'\Big]\Bigr\}\,,
\end{equation}
initialized by the values of $\OPT[\{j\},t,\epsilon]$ for all $j\in[n]$, $t\in T$ and $\epsilon\in E$.
\end{property}

\begin{lemma}~\eqref{eq:comp_d_dpas_recu} solves $\Pcal$ in $\omega(|E|^3\cdot|T|\cdot 2^n)$. 
\end{lemma}
\proof
This proof is essentially the same as the one of Lemma~\ref{lem:Add_D_DPAS_complexity} with the same modifications that for the proof of Lemma~\ref{lemma:Comp_DPAS_complexity}.
\endproof

As for the Additive DPAS, we notice that, with a classical dynamic programming algorithm, the time complexity to solve $\mathcal{P}$ with recurrence~\eqref{eq:comp_dpas_recu} is better than with recurrence~\eqref{eq:comp_d_dpas_recu}. Next, we show that the hybrid algorithm applied to problems satisfying Additive DPAS recurrences can be easily adapted to tackle problems satisfying Composed DPAS recurrences.

\subsection{Hybrid algorithm for Composed DPAS}

The hybrid algorithm for Composed DPAS derives naturally from Algorithm~\ref{alg:QDDPAS}. It amounts to replacing the recurrence~\eqref{eq:add_dpas_recu}, respectivelly~\eqref{eq:add_d_dpas_recu}, by~\eqref{eq:comp_dpas_recu}, respectivelly~\eqref{eq:comp_d_dpas_recu}, resulting in Algorithm~\ref{alg:Comp-QDDPAS}. Eventually, we use Algorithm~\ref{alg:Comp-QDDPAS} as a subroutine to solve Equation~\eqref{eq:def_p_comp_cal}, i.e. to solve the nominal problem $\Pcal$.

\begin{algorithm}[!ht]
\label{alg:Comp-QDDPAS}
\LinesNotNumbered
\KwIn{$\epsilon_0\in E$, auxiliary problem $P'$ satisfying~\eqref{eq:comp_dpas_recu} and~\eqref{eq:comp_d_dpas_recu}}
\KwOut{$\OPT[[n],0,\epsilon_0]$ with high probability}
\caption{Q-DDPAS for Composed DPAS}
\Begin(\textbf{classical part}){
\For{$X \subseteq [n]$ such that $|X| = n/4$, and $t\in T$}{
Compute $\OPT[X,t,\epsilon_0]$ with~\eqref{eq:comp_dpas_recu} and store the results in the QRAM\;
}}
\Begin(\textbf{quantum part}){
Apply Quantum Minimum Finding with~\eqref{eq:comp_d_dpas_recu} to find $\OPT[[n],0,\epsilon_0]$\;
To get values for the Quantum Minimum Finding above (the values $\OPT[J,t,\epsilon]$ for $J\subseteq[n]$ of size $n/2$, $t\in T$ and $\epsilon\in E$), apply Quantum Minimum Finding with~\eqref{eq:comp_d_dpas_recu}\;
To get values for the Quantum Minimum Finding above (the values $\OPT[X,t',\epsilon']$ for $X\subseteq[n]$ of size $n/4$, $t'\in T$ and $\epsilon'\in E$), get them on the QRAM
}
\end{algorithm}

\begin{lemma}
\label{thm:Comp_QDDPAS}
    Let $\epsilon_0 \in E$. The bounded-error algorithm Q-DDPAS (Algorithm~\ref{alg:Comp-QDDPAS}) solves $P'([n],0,\epsilon_0)$ in $\mathcal{O}^*(|E|^2\cdot |T|\cdot 1.754^n)$.
\end{lemma}
Notice that the implementation of this algorithm is slightly different from the one of Algorithm~\ref{alg:QDDPAS}, mainly due to the operation of composition. The details are given in the companion paper~\citep{Grange_CompanionPaper}.

\begin{algorithm}[!ht]
\LinesNotNumbered
\label{alg:meta_comp_QDDPAS}
\KwIn{Auxiliary problem $P'$ satisfying~\eqref{eq:comp_dpas_recu} and~\eqref{eq:comp_d_dpas_recu}}
\KwOut{$\min\limits_{\epsilon\in E} \Bigl\{\epsilon : \OPT[[n],0,\epsilon] < +\infty\Bigr\}$ with high probability}
\caption{Meta-algorithm with subroutine Q-DDPAS for Composed DPAS}
$\epsilon^* \leftarrow +\infty$\;
\For{$\epsilon \in E$}{
Solve $P([n],0,\epsilon)$ with Algorithm~\ref{alg:Comp-QDDPAS}\;
\If{$\OPT[[n],0,\epsilon] < +\infty$ and $\epsilon < \epsilon^*$}{
$\epsilon^* \leftarrow \epsilon$\;
}
}
Return $\epsilon^*$
\end{algorithm}

\begin{theorem}
    The bounded-error Algorithm~\ref{alg:meta_comp_QDDPAS}, with Q-DDPAS as a subroutine, solves $\mathcal{P}$ in $\mathcal{O}^*(|E|^3\cdot |T|\cdot 1.754^n)$.
\end{theorem}

As for the case of Q-DDPAS for Additive DPAS, we can reduce the exponential part of Q-DDPAS complexity for Composed DPAS, by the modification indicated in Observation~\ref{obs:QDDPAS_modif}, thus leading to the following observation.
\begin{observation}
\label{obs:QDDPAS_comp_modif}
    A slight modification of the Q-DDPAS algorithm can reduce the complexity of Algorithm~\ref{alg:meta_comp_QDDPAS} to $\mathcal{O}^*(|E|^3\cdot |T|\cdot 1.728^n)$.
\end{observation}
We illustrate in Subsection~\ref{subsec:scheduling_comp_dpas} the application of Q-DDPAS for Composed DPAS to the problem $1|r_j|\sum w_jU_j$ described in Subsection~\ref{subsec:comp_dpas_example}, together with another similar scheduling problem.

\section{Application to the scheduling literature}
\label{sec:application_scheduling}
In Subsection~\ref{subsec:add_dpas_general} and Subsection~\ref{subsec:comp_dpas_general}, we provided general formulations of problems satisfying Additive and Composed DPAS recurrences. Next, we illustrate these recurrences with several NP-hard single-machine scheduling problems enabling their resolution with our hybrid algorithm Q-DDPAS. The list of problems is non-exhaustive but highlights the structures' specificity of scheduling problems that enable such recurrences. Eventually,  for each problem, we compare in Table~\ref{tab:complexity} the worst-case time complexity of Q-DDPAS with the complexity of the best-known moderate exponential-time exact algorithm. Q-DDPAS improves the exponential-part complexity, sometimes at the cost of an additional pseudo-polynomial factor.

\subsection{Scheduling with deadlines and precedence constraints}
\label{subsec:scheduling_add_dpas}
Single-machine scheduling problems with no constraints, deadline constraints or precedence constraints satisfy the \emph{addition} of optimal values of the problem on sub-instances. We provide next several examples of problems that satisfy Additive DPAS and thus can be solved by Q-DDPAS (Algorithm~\ref{alg:QDDPAS}).

In Subsection~\ref{subsec:add_dpas_example}, we have presented the problem of minimizing the total weighted completion time with deadline constraints ($1|\Tilde{d}_j|\sum_{j}w_jC_j$). The formulation needed the set $T$ to be equal to  $\llbracket0,\mathlarger\sum_{j=1}^n p_j\rrbracket$, hence its resolution with Q-DDPAS is in $\mathcal{O}^*(\sum p_j \cdot 1.728^n)$ according to Observation~\ref{obs:QDDPAS_modif}.
Next, we give two more examples, beginning with the strongly NP-hard scheduling problem with minimization of the total weighted tardiness. Henceforth, we note $p(J) = \sum_{j\in J} p_j$ the sum of processing times of the jobs in $J\subseteq[n]$.

\begin{example}[Minimizing the total weighted tardiness, $1||\sum_j w_jT_j$]
\label{ex:tardiness}
For each job $j\in [n]$, we are given a weight $w_j$, a processing time $p_j$, and a due date $d_j$ that indicates the time after which the job is late. Thus, a job $j$ is late in permutation $\pi$ if its completion time is larger than $d_j$, and we define as $T_j(\pi) = \max(0,C_j(\pi) - d_j)$ its tardiness. Our problem aims at finding the permutation that minimizes the total weighted tardiness, referred to as $1||\sum_j w_jT_j$ in the scheduling literature.
Let $T = \llbracket 0, \sum_{j=1}^n p_j\rrbracket$ be the set of all possible starting times. We define the related problem $P$ of Equation~\eqref{eq:related_problem} as follows: for $J\subseteq \mathbb [n]$ and $t \in T$,
$\Pi(J,t) = S_J,$
and for $\pi \in \Pi(J,t)$:
$$f(\pi,J,t) = \sum_{j\in J}w_j\max(0,C_j(\pi)-d_j+ t)\,,$$
where $\max(0,C_j-d_j + t)$ represents the tardiness of job $j$ for the \emph{effective due date} $(d_j - t)$.
Problem $1||\sum_j w_jT_j$ satisfies both Additive DPAS recurrences. Indeed, Equation~\eqref{eq:add_dpas_recu} is valid with:
$\forall J\subseteq [n], \forall j\in J, \forall t\in T,$
\begin{align*}
    g(J,j,t) = w_j\max(0,p(J) - d_j + t)\,,
\end{align*}
where the computation of $g$ is polynomial (linear).
Moreover, Equation~\eqref{eq:add_d_dpas_recu} is valid for the following functions:
$\forall X\subseteq J\subseteq [n] \mbox{ s.t. } |X| = |J|/2, \forall t\in T,$
\begin{align}
    \tshift(J,X,t) = t + p(X) \qquad\mbox{and}\qquad h(J,X,t) = 0
\end{align}
initialized by, for $j\in [n]$ and $t\in T$, $\OPT[\{j\},t] = w_j\max(0,p_j - d_j + t)\,.$
\end{example}

We consider the scheduling problem with precedence constraints and minimization of the total weighted completion time that is also NP-hard. Conversely to the two previous examples, the set $T$ is reduced to $\{0\}$, and function $h$ translates the potential infeasibility of the concatenation of problem $P$ on two sub-instances.

\begin{example}[Minimizing the total weighted completion time with precedence constraints, $1|prec|\sum_{j}w_jC_j$]
\label{ex:precedence}
We are given, for each job $j\in [n]$, a processing time $p_j$, a weight $w_j$, and a set of precedence constraints $K = \{(i,j) : i \prec j\}$. A pair of jobs $(i,j)$ in $K$ implies that $i$ must precede $j$ in the permutation, namely that $i$ must be processed before $j$. Our problem, denoted by $1|prec|\sum_{j}w_jC_j$, aims at finding the feasible permutation, i.e. that respects the precedence constraints, that minimizes the total weighted completion time.
Let be $T = \{0\}$. Here, an instance of the problem $P$ of Equation~\eqref{eq:related_problem} under consideration is only indexed by the chosen subset of $[n]$. Thus, we consider the problem $P$  as follows: for $J\subseteq \mathbb [n]$,
$\Pi(J,0) = \{\pi \in S_J\,|\, \pi \text{ respects } K\}\,,$
and for $\pi \in \Pi(J,0)$:
$f(\pi,J,0) = \sum_{j\in J} w_jC_j(\pi) \,.$
Our problem $1|prec|\sum_{j}w_jC_j$ satisfies both Additive DPAS recurrences. Indeed, Equation~\eqref{eq:add_dpas_recu} is valid for:
$$\forall J\subseteq [n], \forall j\in J,~~
    g(J,j,0) =\left\{
     \begin{aligned}
      & +\infty &\text{if } \exists (j,k)\in E | k\in J\\
      & w_jp(J)
      &\text{otherwise}
    \end{aligned}
  \right.\,,$$
where the computation of $g$ is polynomial (quadratic). 
This problem $P$ also satisfies~\eqref{eq:add_d_dpas_recu}. Indeed, Equation~\eqref{eq:add_d_dpas_recu} is valid for the following functions:
$\forall X\subseteq J\subseteq [n] $ such that $ |X| = |J|/2,$
\begin{align*}
    \tshift(J,X,0) = 0 \qquad\mbox{and}\qquad
    h(J,X,0) = \left\{
     \begin{aligned}
      & +\infty &\text{if } \exists (j,k)\in E | j\in J\setminus X \text{ and } k\in X\\
      & p(X)\cdot\sum_{j\in J\setminus X}w_j &\text{otherwise}
    \end{aligned}
  \right.
\end{align*}
where the computation of $h$ is polynomial (quadratic). 
The initialization is, for $j\in [n]$, $\OPT[\{j\},0] = w_jp_j\,.$
\end{example}

The three NP-hard scheduling problems examples described above can be solved with Q-DDPAS for Additive DPAS (Algorithm~\ref{alg:QDDPAS}).
We illustrate in Table~\ref{tab:complexity_add_dpas} the worst-case time complexities of solving them with Q-DDPAS and compare them with the complexities of the best-known exact classical algorithms. Q-DDPAS improves the complexity of the exponent but sometimes at the cost of a pseudo-polynomial factor. 

\begin{table}[ht]
\small
\begin{center}
\begin{tabular}{||c c c||}
 \hline
 Problem & Q-DDPAS for Additive DPAS  & Best classical algorithm\\ [1ex] 
 \hline\hline
 $1|\Tilde{d}_j|\sum w_jC_j$ & $\mathcal{O}^* \left(\sum p_j\cdot1.728^n\right)$ & $\mathcal{O}^*(2^n)$~\citep{TKindt}\\ [1ex]
 \hline
 $1||\sum w_jT_j$ & $\mathcal{O}^* \left(\sum p_j\cdot1.728^n\right)$ & $\mathcal{O}^*(2^n)$~\citep{TKindt}\\ [1ex]
 \hline
 $1|prec|\sum w_jC_j$ & $\mathcal{O}^* \left(1.728^n\right)$ & $\mathcal{O}^*((2-\epsilon)^n)$, for small $\epsilon$~\citep{Cygan}\\ [1ex]
 \hline
\end{tabular}
\end{center}
\caption{Comparison of complexities between Q-DDPAS and the best-known classical algorithm for some scheduling problems satisfying~\eqref{eq:add_dpas_recu} and~\eqref{eq:add_d_dpas_recu}.}
 \label{tab:complexity_add_dpas}
 \end{table}

\subsection{Scheduling with release date constraints}
\label{subsec:scheduling_comp_dpas}
Single-machine scheduling problems with release date constraints do not satisfy the \emph{addition} of optimal values of the problem on sub-instances but enable the \emph{composition} of them. We illustrate this notion with two examples of problems that satisfy Composed DPAS and thus can be solved by Q-DDPAS (Algorithm~\ref{alg:Comp-QDDPAS}).

We have presented in Subsection~\ref{subsec:comp_dpas_example} an example that is the problem of minimizing the weighted number of late jobs with release date constraints ($1|r_j|\sum w_jU_j$). We have shown that the two sets to define the auxiliary problem are $E = \llbracket0,\sum_{j=1}^n w_j\rrbracket$ and $T = \llbracket0,\sum_{j=1}^n p_j\rrbracket\cup \{+\infty\}$. Thus, Q-DDPAS solves this problem in $\mathcal{O}^* \left((\sum w_j)^3\cdot \sum p_j \cdot 1.728^n\right)$ according to Observation~\ref{obs:QDDPAS_comp_modif}. Next, we present another example which is the strongly NP-hard problem of minimizing the total weighted completion time with release date constraints.

\begin{example}[Minimizing the total weighted completion time with release date constraints, $1|r_j|\sum w_jC_j$]
Each job $j\in [n]$ has a weight $w_j$, a processing time $p_j$, and a release date $r_j$. This problem aims at finding the feasible permutation, namely where each job starts after its release date, for which the total weighted completion time is minimal. 
Let $T =\llbracket0,\sum_{j=1}^n p_j\rrbracket\cup \{+\infty\}$ and $E = \llbracket0,\sum_{j=1}^n w_j \cdot \sum_{j=1}^n p_j\rrbracket$. For a given $\epsilon\in E$, we consider the problem $P'$ of Equation~\eqref{eq:related_auxiliary_problem} as follows: $\forall J\subseteq[n], t\in T,$ 
    \begin{equation*}
        P'(J,t,\epsilon) : ~~ \min_{\pi\in \Pi'(J,t,\epsilon)} C_\textup{max}(\pi)\,,
    \end{equation*}
    where $C_\text{max}$ is the makespan, and 
    \begin{equation*}
        \Pi'(J,t,\epsilon) = \{\pi\in S_J : C_j(\pi)\geq \max(t,r_j) + p_j \mbox{ and } \sum_{j\in J} w_jC_j(\pi) = \epsilon\}\,,
    \end{equation*}
    where $C_j$ is the completion time of job $j$.
Problem $P'$ satisfies the two Composed DPAS recurrences~\eqref{eq:comp_dpas_recu} and~\eqref{eq:comp_d_dpas_recu}. The initialization of the recurrences is, for $j\in[n]$, $t\in T$ and $\epsilon\in E$,
\begin{equation*}
\OPT[\{j\},t,\epsilon] =
    \begin{cases}
        C_j := \max(t,r_j)+p_j,~~\text{ if } \epsilon = w_jC_j\\
        +\infty ,~~\text{otherwise}
    \end{cases}
\end{equation*}
\end{example} 

We synthesize in Table~\ref{tab:complexity_comp_dpas} the worst-case time complexities achieved by Q-DDPAS on the examples of scheduling problems satisfying the Composed DPAS recurrences. We compare them with the best-known classical complexities for exact algorithms. The latter comes from the algorithm of Inclusion-Exclusion designed by~\cite{Ploton}, which provides a generic method to solve such problems. We observe that Q-DDPAS improves the exponential part of the complexity, at a cost of a higher degree for the pseudo-polynomial factor. 

\begin{table}[ht]
\small
\begin{center}
\begin{tabular}{||c c c||}
 \hline
 Problem & Q-DDPAS for Composed DPAS & Best classical algorithm\\ [1ex] 
 \hline\hline
  $1|r_j|\sum w_jU_j$ & $\mathcal{O}^* \left((\sum w_j)^3\cdot \sum p_j \cdot 1.728^n\right)$ & $\mathcal{O}^*(\sum w_j\cdot \sum p_j\cdot2^n)\,,$~\citep{Ploton}\\ [1ex]
 \hline
 $1|r_j|\sum w_jC_j$ & $\mathcal{O}^* \left((\sum w_j)^3\cdot (\sum p_j)^4 \cdot 1.728^n\right)$ & $\mathcal{O}^*(\sum w_j\cdot (\sum p_j)^2\cdot2^n)\,,$~\citep{Ploton}\\ [1ex]
 \hline
\end{tabular}
\end{center}
\caption{Comparison of complexities between Q-DDPAS and the best-known classical algorithm for some scheduling problems satisfying~\eqref{eq:comp_dpas_recu} and~\eqref{eq:comp_d_dpas_recu}.}
\label{tab:complexity_comp_dpas}
\end{table}

\section{Decision-based DPAS}
\label{sec:dec_dpas}
We saw in the previous section that the recurrence to solve $\mathcal{P}$ can be applied to a minimization problem, possibly involving an auxiliary problem. Sometimes, the recurrence does not apply directly to a minimization problem but to a \emph{decision} problem. This is the case of the 3-machine flowshop problem. In this section, we adapt the hybrid algorithm Q-DDPAS to solve this problem. Notice that it easily generalizes to the $m$-flowshop problem, for $m\geq 4$.

\subsection{3-machine flowshop and dynamic programming}
We consider the permutation flowshop problem on 3 machines for $n$ jobs with minimizing the makespan as the objective function. This strongly NP-hard problem is often referred to as $F3||C_\text{max}$ in the literature, as mentioned by~\cite{Tkindt_3FlowShop}. Each job $j\in[n]$ consists of 3 operations $O_{ij}$ for $i\in[3]$, each operation being processed on the $i$-th machine. We note $p_{ij}$ the processing time of operation $O_{ij}$. Each machine performs at most one operation at a time. For each job $j$, operations must be processed in the specific order $O_{1j}$, $O_{2j}$, $O_{3j}$: the first operation gets processed on the first machine, then the second operation gets processed on second machine (as soon as the first operation is finished and the second machine is available), and eventually the third operation gets processed on the third machine (as soon as the second operation is finished and the third machine is available). Thus, only the processing order of the jobs has to be decided, implying that a solution is entirely described by the permutation of jobs on the first machine. Thus, the problem can be formulated as
\begin{equation}
\label{eq:flowshop}
\min\limits_{\pi \in S_{[n]}} C_\text{max}(\pi)\,,
\end{equation}
where $C_\text{max}$ is the maximum completion time, namely the completion time of the last job processed on the last machine (third machine). Because the two techniques presented so far do not apply to~\eqref{eq:flowshop}, we present an alternative approach involving the decision counterpart of the above optimization problem. 

We introduce below a decision problem for deriving the recurrences. For that, we define the bounded set $$T = \left\llbracket0,\mathlarger\sum_{j\in[n], i\in[3]} p_{ij}\right\rrbracket\subseteq \Z\,.$$ 

\begin{definition}[Decision problem]
\label{def:dec_problems}
For $J\subseteq[n]$, $\bvect = (\beta_2, \beta_3) \in T^2$ and $ \evect = (\epsilon_2, \epsilon_3)\in T^2$, we define the decision problem $D(J, \bvect,\evect)$ on a sub-instance associated with jobs in $J$ as the following question: 
``Does there exist a permutation $\pi\in S_J$ such that, for $i \in \{2,3\}$, $b_i(\pi) \geq \beta_i\,,$ and $e_i(\pi) \leq \epsilon_i\,?"$,
where $b_i(\pi)$, respectively $e_i(\pi)$, denotes the time at which the first operation begins, respectively the last operation ends, on the $i$-th machine. 
\end{definition}

In other words, problem $\SAT(J,\bvect, \evect)$ asks whether or not there exists a feasible permutation with jobs in $J$ such that it $\emph{holds}$ between the two $\emph{temporal fronts}$ $\bvect$ and $\evect$.
Notice that it is not necessary to impose any beginning and ending time for the first machine ($i=1$). Indeed, the problem is time-invariant, thus we can always consider that the scheduling problem starts at time 0, and that the total completion time on the first machine is known and equal to the sum of processing times of the scheduled jobs. 
Notice that the number of parameters is four for the 3-machine flowshop, but generalizes to $2(m-1)$ parameters for the $m$-machine flowshop.

With these notations, $\mathcal{P}$ can be cast as follows:
\begin{equation}
\label{eq:def_p_dec}
    \mathcal{P} : \hspace{6mm} \min_{c\in T} \Bigl\{c : \SAT[[n],(0,0),(c,c)] = \text{True}\Bigr\}\,.
\end{equation}
The decision problem $D$ satisfies both  recurrences~\eqref{eq:dec_dpas} and~\eqref{eq:dec_d_dpas} below. 
\begin{property}[Decision DPAS]
For all $J\subseteq [n]$ of even cardinality, $\bvect\in T^2$ and $\evect\in T^2$,
    \begin{equation}
    \tag{Dec-DPAS}
    \label{eq:dec_dpas}
        \SAT[J,\bvect, \evect] = \bigvee\limits_{X\subseteq J : |X| = |J|/2,\atop \tvect\in [\bvect, \evect]} \left(\SAT[\{j\},\bvect, \tvect] \land \SAT[J\setminus \{j\},\tvect \ominus p_{1j}, \evect\ominus p_{1j}]\right)\,,
    \end{equation}
where $\tvect\in [\bvect, \evect]$ means that the $i$-th coordinate of $\tvect$ is between the $i$-th coordinates of $\bvect$ and $\evect$, and where operation $\Vec{v} \ominus c$, for a vector $\Vec{v}$ and a constant $c$, subtracts $c$ to each coordinate of $\Vec{v}$.
\end{property}
This latter recurrence enables $\mathcal{P}$ to be solved by a classical dynamic programming algorithm.

\begin{lemma}
\label{lem:DecDPAS_complexity}
~\eqref{eq:dec_dpas} solves $\mathcal{P}$ in $\mathcal{O}^*(|T|^4\cdot 2^n)$.
\end{lemma}
\proof
First, we can show that, for a given $\bvect_0,\evect_0\in T^2$,~\eqref{eq:dec_dpas} solves $\SAT([n],\bvect_0,\evect_0)$ in $\mathcal{O}^*(|T|^4\cdot 2^n)$. This is essentially the same lines of the proof as in Lemma~\ref{lemma:Add_DPAS_complexity}.
Second, to solve $\mathcal{P}$, we make a dichotomic search over $T$ to find the minimum $c\in T$ such that $\SAT([n],(0,c),(0,c))$ is True according to Equation~\eqref{eq:def_p_dec}. Thus,~\eqref{eq:dec_dpas} is called $\log_2(|T|)$ times. Because $|T|=\sum p_{ij}$ is a pseudo-polynomial term of the instance, the total complexity is $\mathcal{O}^*(\log_2(|T|)\cdot |T|^4\cdot 2^n) = \mathcal{O}^*(|T|^4\cdot 2^n)\,.$
\endproof

\begin{property}[Decision Dichotomic DPAS]
For all $J\subseteq [n]$ of even cardinality, $\bvect\in T^2$ and $\evect\in T^2$,
    \begin{equation}
    \tag{Dec-D-DPAS}
    \label{eq:dec_d_dpas}
        \SAT[J,\bvect, \evect] = \bigvee\limits_{X\subseteq J : |X| = |J|/2,\atop \tvect\in [\bvect, \evect]} \left(\SAT[X,\bvect, \tvect] \land \SAT[J\setminus X,\tvect \ominus \sum_{j\in X}p_{1j}, \evect\ominus \sum_{j\in X}p_{1j}]\right)\,.
    \end{equation}
\end{property}

\begin{lemma}
 ~\eqref{eq:dec_d_dpas} solves $\mathcal{P}$ in $\omega(|T|^4\cdot 2^n)$.
\end{lemma}

\proof
This proof is similar to the proof of Lemma~\ref{lem:Add_D_DPAS_complexity}, with the argument that a dichotomic search is polynomial in the size of the instance as in the proof of Lemma~\ref{lem:DecDPAS_complexity}.
\endproof
Once again, we observe that recurrence~\eqref{eq:dec_dpas} outperforms recurrence~\eqref{eq:dec_d_dpas} to solve by classical dynamic programming our problem $\mathcal{P}$. In the next subsection, we describe how we adapt Q-DDPAS to take advantage of those two recurrences to solve the 3-machine flowshop problem.

\subsection{Hybrid algorithm for Decision-based DPAS}
We call Q-Dec-DDPAS the adapted decision version of Q-DDPAS. The main difference is that instead of searching for a minimum value in a set in recurrence~\eqref{eq:add_d_dpas_recu} or~\eqref{eq:comp_d_dpas_recu}, we search for a True value in a set in recurrence~\eqref{eq:dec_d_dpas}. Thus, it essentially amounts to replacing Quantum Minimum Finding with the algorithm of~\cite{Boyer98} specified below, which extends Grover Search~\citep{Grover}. 

\begin{definition}[Grover Search Extension~\citep{Boyer98}]
Let $f:[n]\rightarrow \{0,1\}$ be a function. Grover Search Extension computes with high probability the logical OR of all the $f$ values and the corresponding antecedent(s) $x\in [n]$ such that $f(x) = 1$. The complexity of Grover Search Extension is $\mathcal{O}\left(\sqrt{n}\cdot C_f(n)\right)$, where $\mathcal{O}(C_f(n))$ is the complexity of computing a value of $f$.
\end{definition}

Notice that we cannot use Grover Search because we do not know the number of $x$ such that $f(x) = 1$. The generalization of~\cite{Boyer98} enables us to deal with an unknown number of solutions while keeping the same complexity of Grover Search. Moreover, if there are $t\in\mathbb{N}^*$ solutions, the complexity is $\mathcal{O}\left(\sqrt{n/t}\cdot C_f(n)\right)$ but, having no bounds on $t$ whenever we call Grover Search Extension, we omit it in the complexity.

\begin{algorithm}[!ht]
\label{alg:Dec-QDDPAS}
\LinesNotNumbered
\KwIn{$\bvect_0, \evect_0\in T^2$, decision problem $D$ satisfying~\eqref{eq:dec_dpas} and~\eqref{eq:dec_d_dpas}}
\KwOut{$\SAT[[n],\bvect_0,\evect_0]$ with high probability}
\caption{Q-Dec-DDPAS for 3-machine flowshop}
\Begin(\textbf{classical part}){
\For{$X \subseteq [n] : |X| = n/4$ and $\bvect, \evect\in T^2$}{
Compute $\SAT[X,\bvect, \evect]$ with~\eqref{eq:dec_dpas} and store the results in the QRAM\;
}}
\Begin(\textbf{quantum part}){
Apply Grover Search Extension with~\eqref{eq:dec_d_dpas} to find $\SAT[[n],\bvect_0,\evect_0]$\;
To get values for the Grover Search Extension above (the values $\SAT[J,\bvect,\evect]$ for $J\subseteq[n]$ of size $n/2$ and $\bvect,\evect\in T$), apply Grover Search Extension with~\eqref{eq:dec_d_dpas}\;
To get values for Grover Search Extension above (the values $\SAT[X,\bvect',\evect']$ for $X\subseteq[n]$ of size $n/4$ and $\bvect',\evect'\in T$), get them on the QRAM\;
}
\end{algorithm}

\begin{restatable}{lemma}{QDecDDPAScomplexity}
   Let $\bvect_0, \evect_0\in T^2$. The bounded-error algorithm Q-Dec-DDPAS (Algorithm~\ref{alg:Dec-QDDPAS}) solves $D([n],\bvect_0, \evect_0)$ in $\mathcal{O}^*((\sum p_{ij})^4\cdot1.754^n)$. 
\end{restatable}

The proof is essentially the same as for Theorem~\ref{thm:QDDPAS}, detailed in Supplementary Materials. All the details of correctness and low-level implementation are given in our companion paper~\citep{Grange_CompanionPaper}.

\begin{algorithm}[!ht]
\LinesNotNumbered
\label{alg:meta_dec_QDDPAS}
\KwIn{3-machine flowshop}
\KwOut{Minimum makespan with high probability}
\caption{Meta-algorithm with subroutine Q-Dec-DDPAS for the 3-machine flowshop}
$c^* \leftarrow +\infty$\;
\For{$c \in T$}{
Solve $\SAT([n],(0,0),(c,c))$ with Algorithm~\ref{alg:Dec-QDDPAS}\;
\If{$D[[n],(0,0),(c,c)]$ and $c < c^*$}{
$c^* \leftarrow c$\;
}
}
Return $c^*$
\end{algorithm}

\begin{theorem}
    The bounded-error Algorithm~\ref{alg:meta_dec_QDDPAS} solves the 3-machine flowshop in $\mathcal{O}^*((\sum p_{ij})^4\cdot 1.754^n)$ with high probability. 
\end{theorem}

Once again, as mentioned in Observation~\ref{obs:QDDPAS_modif}, the complexity can be reduced thanks to a slight modification on the Q-Dec-DDPAS that constitutes the subroutine, thus leading to the following observation.
\begin{observation}
A slight modification of Algorithm~\ref{alg:meta_dec_QDDPAS} reduces the complexity of solving the 3-machine flowshop in $\mathcal{O}^*((\sum p_{ij})^4\cdot 1.728^n)$ with high probability. 
\end{observation}

This new method improves the best-known classical algorithm that is in $\mathcal{O}^*(3^n)$ or in $\mathcal{O}^*(M\cdot 2^n)$ if there exists a constant $M$ such that $p_{ij} \leq M$, for all $i\in [3], j\in[n]$, presented by~\cite{Tkindt_3FlowShop} and~\cite{Ploton_flowshop}. Hybrid quantum-classical bounded-error Algorithm~\ref{alg:meta_dec_QDDPAS} reduces the exponential part of the time complexity at the cost of a pseudo-polynomial factor. For most cases, this factor is negligible because the numerical values of 3-machine flowshop instances are small compared to the exponential part value. However, we present in the next subsection a way to dispose of this factor with an approximation scheme. 

It is worth noting that the previous algorithm easily generalizes to the $m$-machine flowshop problem. Indeed, the only difference is the description of the \emph{temporal front} that necessitates $2(m-1)$ parameters.
\begin{observation}
The bounded-error Algorithm~\ref{alg:meta_dec_QDDPAS} generalizes to solve the $m$-machine flowshop in $\mathcal{O}^*((\sum p_{ij})^{2(m-1)}\cdot 1.728^n)$ with high probability. 
\end{observation}

Notice that~\cite{Ploton_flowshop} present a classical resolution for the $m$-machine flowshop by Inclusion-Exclusion in $\mathcal{O}^*((\sum p_{ij})^{m}\cdot 2^n)$.

\subsection{Approximation scheme for the 3-machine flowshop}
We present an approximation scheme for the 3-machine flowshop problem that trades the pseudo-polynomial factor in the complexity of Q-Dec-DDPAS and the optimality of the algorithm for a polynomial factor in $\frac{1}{\epsilon}$ and an approximation factor of $(1+\epsilon)$. In other words, we provide Algorithm~\ref{alg:approx_scheme} that finds a solution in time $\mathcal{O}^*\left(\frac{1}{\epsilon^3}\cdot 1.728^n\right)$ for which the makespan is not greater than (1+$\epsilon$) times the optimal makespan. The latter point denotes that this is an $\epsilon$-approximation scheme. Our algorithm belongs to the class of moderate exponential-time approximation algorithms. Notice that the 3-machine flowshop problem does not admit an FPTAS (fully polynomial-time approximation scheme) because it is strongly NP-hard, meaning that no $\epsilon$-approximation algorithm exists to solve the 3-machine flowshop in time $\mathcal{O}\left(poly(n,\frac{1}{\epsilon})\right)$ unless P = NP~\citep{VaziraniApprox}. In comparison,~\cite{Hall} provides for the $m$-machine flowshop problem an FPT-AS (fixed-parameter tractable approximation scheme), namely an $\epsilon$-approximation algorithm that runs in time $\mathcal{O}(f(\epsilon,\kappa)\cdot poly(n))$ for $\kappa$ a fixed parameter of the instance and $f$ a computable function.~\cite{Hall} choose $\kappa$ to be the number of machines of the flowshop, leading to an FPT-AS that runs in time $\mathcal{O}\left(n^{3.5}\cdot (\frac{m}{\epsilon})^\frac{m^4}{\epsilon^2}\right)$. In our case, we consider the case $m=3$.

\begin{algorithm}[!ht]
\label{alg:approx_scheme}
\KwIn{$\epsilon > 0$, 3-machine flowshop on $n$ jobs with processing times $\{p_{ij} : i\in [3],j\in[n]\}$}
\KwOut{solution at most $1+\epsilon$ times the optimal solution}
\caption{Hybrid approximation scheme for the 3-machine flowshop}
$P = \max\limits_{i\in [3],j\in[n]} \{p_{ij}\}$\;
$K = \frac{\epsilon P}{n+2}$\;
\For{$i\in[3], j\in [n]$}{
$p'_{ij} = \lceil \frac{p_{ij}}{K}\rceil$\;
}
Solve the 3-machine flowshop on $n$ jobs with new processing times $\{p'_{ij} : i\in [3],j\in[n]\}$ with Algorithm~\ref{alg:meta_dec_QDDPAS} that outputs permutation $\pi'$\;
Return $\pi'$
\end{algorithm}

\begin{lemma}
\label{lem:epsilon_approx}
    Let $\pi^*$ be an optimal solution of the 3-machine flowshop problem, for the processing times $\{p_{ij} : i\in [3],j\in[n]\}$. Let $\pi'$ be the output of Algorithm~\ref{alg:approx_scheme}. We have 
    \begin{equation*}
        \Cmax(\pi') \leq (1+\epsilon)\cdot\Cmax(\pi^*)\,.
    \end{equation*}
\end{lemma}
Next, we introduce two observations necessary to prove Lemma~\ref{lem:epsilon_approx}. The proofs are omitted because of their simplicity. 
\begin{observation}
\label{prop:multiplication_makespan}
    Let $\pi$ be a permutation and let be $\alpha\in \R^*_+$. We note $\Cmax(\pi)$ the makespan of $\pi$ of the 3-machine flowshop for processing times $\{p_{ij} : i\in [3],j\in[n]\}$.
    We note $\Cmax'(\pi)$ the makespan of $\pi$ of the 3-machine flowshop for processing times $\{p'_{ij} : i\in [3],j\in[n]\}$ such that $p'_{ij} := \alpha p_{ij}$ for all $i,j$. Then, 
    $\Cmax'(\pi) = \alpha\Cmax(\pi)\,.$
    Notice that for $p'_{ij} \leq  \alpha p_{ij}$, we have $\Cmax'(\pi)\leq \alpha \Cmax(\pi)$ even if the critical path in $\pi$ may differ to obtain $\Cmax$ and $\Cmax'$.
\end{observation}

\begin{observation}
\label{prop:addition_makespan}
    Let $\pi$ be a permutation and let $\beta\in\R$ such that $\beta \geq -\min\limits_{i\in [3],j\in[n]} \{p_{ij}\}$. We note $\Cmax(\pi)$ the makespan of $\pi$ of the 3-machine flowshop for processing times $\{p_{ij} : i\in [3],j\in[n]\}$.
    We note $\Cmax''(\pi)$ the makespan of $\pi$ of the 3-machine flowshop for processing times $\{p''_{ij} : i\in [3],j\in[n]\}$ such that $p''_{ij} := p_{ij} + \beta$ for all $i\in[3],j\in[n]$. Then, 
    $\Cmax''(\pi) \leq \Cmax(\pi) + \beta(n+2)\,.$
    Notice that for $p''_{ij} \leq  p_{ij} + \beta$, we still have $\Cmax''(\pi)\leq \Cmax(\pi) + \beta(n+2)$ even if the critical path in $\pi$ may differ to obtain $\Cmax$ and $\Cmax''$.
\end{observation}

\proof[Proof of Lemma~\ref{lem:epsilon_approx}]
Let be $\epsilon >0$. The new processing times considered $p'_{ij} := \lceil \frac{p_{ij}}{K}\rceil$ imply that $\frac{p_{ij}}{K} \leq p'_{ij} < \frac{p_{ij}}{K} + 1$. We note $\Cmax'$ the makespan of the new problem, i.e. the 3-machine flowshop problem with processing times $\{p'_{ij} : i\in [3],j\in[n]\}$.

On the one hand, we have $p'_{ij} < \frac{p_{ij}}{K} + 1$, for all $i\in [3], j\in[n]$. Thus, according to Observations~\ref{prop:multiplication_makespan} and~\ref{prop:addition_makespan} considering the optimal permutation $\pi^*$,
$\Cmax'(\pi^*) \leq \frac{\Cmax(\pi^*)}{K} + n + 2\,,$
namely, because $K>0$,
\begin{equation}
\label{eq:approx_ineg}
    K\Cmax'(\pi^*) \leq \Cmax(\pi^*) + K(n + 2)\,.
\end{equation}

On the other hand, we have $\frac{p_{ij}}{K} \leq p'_{ij}$. Thus, according to Observation~\ref{prop:multiplication_makespan} considering the output permutation $\pi'$ of Algorithm~\ref{alg:approx_scheme},
$\frac{\Cmax(\pi')}{K} \leq \Cmax'(\pi')\,,$
namely, because $K>0$,
\begin{align}
    \Cmax(\pi') &\leq K\Cmax'(\pi')\leq K\Cmax'(\pi^*) \label{eq:approx_1}\\
    &\leq  \Cmax(\pi^*) + K(n+2) = \Cmax(\pi^*) + \epsilon P \label{eq:approx_2}\\
    &\leq  \Cmax(\pi^*) + \epsilon\Cmax(\pi^*) = (1+\epsilon)\Cmax(\pi^*)\,, \label{eq:approx_3}
\end{align}
where~\eqref{eq:approx_1} comes from the fact that $\pi'$ is the optimal solution for makespan $\Cmax'$,~\eqref{eq:approx_2} results from Equation~\eqref{eq:approx_ineg}, and~\eqref{eq:approx_3} is true because the makespan is always larger than $P = \max\limits_{i\in [3],j\in[n]} \{p_{ij}\}$.
\endproof

\begin{theorem}
    Algorithm~\ref{alg:approx_scheme} is an approximation scheme for the 3-machine flowshop problem and outputs a solution whose makespan it at most $(1+\epsilon)$ times the optimal value in time $\mathcal{O}^*\left(\frac{1}{\epsilon^{3}}\cdot1.728^n\right)\,.$
\end{theorem}

\proof
First, according to Lemma~\ref{lem:epsilon_approx}, Algorithm~\ref{alg:approx_scheme} outputs a solution whose makespan it at most $(1+\epsilon)$ times the optimal value. 
Second, Algorithm~\ref{alg:meta_dec_QDDPAS} solves the new problem in time $\mathcal{O}^*((\sumparam')^{4}\cdot 1.728^n) = \mathcal{O}^*(\frac{1}{\epsilon^{4}}\cdot1.728^n)$. Indeed, 
\begin{align*}
    \sumparam' \leq \sum\left(\frac{p_{ij}}{K} + 1\right) = \frac{1}{K}\sumparam + 3n
    \leq \frac{1}{K}\cdot 3nP + 3n
    = \frac{3n(n+2)}{\epsilon} + 3n\,.   
\end{align*}
Thus, $\sumparam' \leq poly(n, \frac{1}{\epsilon})$.
\endproof

\section{Conclusion}
In this work, we propose generalized dynamic programming recurrences for NP-hard scheduling problems to solve a broad class of such problems with our hybrid algorithm Q-DDPAS which is an adapted version of the quantum-classical algorithm of~\cite{Ambainis}. Q-DDPAS provides a quantum speed-up to their exact resolution. Specifically, our hybrid algorithm reduces the best-known classical time complexity, often equal to $\mathcal{O}^*(2^n)$ for single-machine problems and $\mathcal{O}^*(3^n)$ for the 3-machine flowshop, to $\mathcal{O}^*(1.728^n)$, sometimes at the cost of an additional pseudo-polynomial factor as summarized in Table~\ref{tab:complexity}.
Future work will be dedicated to widening the range of problems for which we can achieve a quantum speed-up. This will necessitate finding NP-hard problems for which the description of a solution is a permutation, and that satisfy a dynamic programming property admitting a dichotomic version, relying either on an optimization or a decision problem. A first lead is to consider the $3$-machine jobshop problem. Indeed, a promising description of a solution of this problem is the Bierwith vector~\citep{Bierwirth95}, a vector encoding a solution as a permutation with repetition of size $3n$, where $n$ is the number of jobs.

\paragraph{Acknowledgements.} This work has been partially financed by the ANRT through the PhD number 2021/0281 with CIFRE funds.

\bibliographystyle{apalike}
\bibliography{ref}

\begin{thebibliography}{}

\bibitem[Ambainis et~al., 2019]{Ambainis}
Ambainis, A., Balodis, K., Iraids, J., Kokainis, M., Pr{\=u}sis, K., and Vihrovs, J. (2019).
\newblock Quantum speedups for exponential-time dynamic programming algorithms.
\newblock In {\em Proceedings of the Thirtieth Annual ACM-SIAM Symposium on Discrete Algorithms}, pages 1783--1793. SIAM.

\bibitem[Bernstein and Vazirani, 1993]{Bernstein93}
Bernstein, E. and Vazirani, U. (1993).
\newblock Quantum complexity theory.
\newblock In {\em Proceedings of the twenty-fifth annual ACM symposium on Theory of computing}, pages 11--20.

\bibitem[Bierwirth, 1995]{Bierwirth95}
Bierwirth, C. (1995).
\newblock A generalized permutation approach to job shop scheduling with genetic algorithms.
\newblock {\em Operations-Research-Spektrum}, 17(2):87--92.

\bibitem[Boyer et~al., 1998]{Boyer98}
Boyer, M., Brassard, G., H{\o}yer, P., and Tapp, A. (1998).
\newblock Tight bounds on quantum searching.
\newblock {\em Fortschritte der Physik: Progress of Physics}, 46(4-5):493--505.

\bibitem[Cerezo et~al., 2021]{Cerezo}
Cerezo, M., Arrasmith, A., Babbush, R., Benjamin, S.~C., Endo, S., Fujii, K., McClean, J.~R., Mitarai, K., Yuan, X., Cincio, L., et~al. (2021).
\newblock Variational quantum algorithms.
\newblock {\em Nature Reviews Physics}, 3(9):625--644.

\bibitem[Cygan et~al., 2014]{Cygan}
Cygan, M., Pilipczuk, M., Pilipczuk, M., and Wojtaszczyk, J.~O. (2014).
\newblock Scheduling partially ordered jobs faster than $2^n$.
\newblock {\em Algorithmica}, 68:692--714.

\bibitem[Dürr and H{\o}yer, 1996]{DurrHoyer}
Dürr, C. and H{\o}yer, P. (1996).
\newblock A quantum algorithm for finding the minimum.
\newblock {\em arXiv preprint quant-ph/9607014}.

\bibitem[Farhi et~al., 2014]{Farhi}
Farhi, E., Goldstone, J., and Gutmann, S. (2014).
\newblock A quantum approximate optimization algorithm.
\newblock {\em arXiv preprint arXiv:1411.4028}.

\bibitem[Giovannetti et~al., 2008]{QRAM}
Giovannetti, V., Lloyd, S., and Maccone, L. (2008).
\newblock Quantum random access memory.
\newblock {\em Physical review letters}, 100(16):160501.

\bibitem[Graham et~al., 1979]{Graham}
Graham, R.~L., Lawler, E.~L., Lenstra, J.~K., and Kan, A.~R. (1979).
\newblock Optimization and approximation in deterministic sequencing and scheduling: a survey.
\newblock In {\em Annals of discrete mathematics}, volume~5, pages 287--326. Elsevier.

\bibitem[Grange et~al., 2023a]{GrangeGecco}
Grange, C., Bourreau, E., Poss, M., and t'Kindt, V. (2023a).
\newblock Quantum speed-ups for single-machine scheduling problems.
\newblock In {\em Proceedings of the Companion Conference on Genetic and Evolutionary Computation}, pages 2224--2231.

\bibitem[Grange et~al., 2023b]{Grange}
Grange, C., Poss, M., and Bourreau, E. (2023b).
\newblock An introduction to variational quantum algorithms for combinatorial optimization problems.
\newblock {\em 4OR}, pages 1--41.

\bibitem[Grange et~al., 2024]{Grange_CompanionPaper}
Grange, C., Poss, M., Bourreau, E., t'Kindt, V., and Ploton, O. (2024).
\newblock {Companion Paper: Moderate Exponential-time Quantum Dynamic Programming Across the Subsets for Scheduling Problems}.
\newblock HAL preprint, https://hal.science/hal-04296238.

\bibitem[Grover, 1996]{Grover}
Grover, L.~K. (1996).
\newblock A fast quantum mechanical algorithm for database search.
\newblock In {\em Proceedings of the twenty-eighth annual ACM symposium on Theory of computing}, pages 212--219.

\bibitem[Hall, 1998]{Hall}
Hall, L.~A. (1998).
\newblock Approximability of flow shop scheduling.
\newblock {\em Mathematical Programming}, 82(1-2):175--190.

\bibitem[Held and Karp, 1970]{HeldKarp}
Held, M. and Karp, R.~M. (1970).
\newblock The traveling-salesman problem and minimum spanning trees.
\newblock {\em Operations Research}, 18(6):1138--1162.

\bibitem[Kurowski et~al., 2023]{JobShop}
Kurowski, K., Pecyna, T., Slysz, M., Rozycki, R., Waligora, G., and Weglarz, J. (2023).
\newblock Application of quantum approximate optimization algorithm to job shop scheduling problem.
\newblock {\em European Journal of Operational Research}, 310(2):518--528.

\bibitem[Lawler, 1990]{Lawler}
Lawler, E.~L. (1990).
\newblock A dynamic programming algorithm for preemptive scheduling of a single machine to minimize the number of late jobs.
\newblock {\em Annals of Operations Research}, 26:125--133.

\bibitem[Miyamoto et~al., 2020]{SteinerTree}
Miyamoto, M., Iwamura, M., Kise, K., and Gall, F.~L. (2020).
\newblock Quantum speedup for the minimum steiner tree problem.
\newblock In {\em COCOON 2020, Atlanta, GA, USA, August 29--31, 2020, Proceedings}, pages 234--245. Springer.

\bibitem[Nannicini, 2019]{Nannicini_MAX3SAT}
Nannicini, G. (2019).
\newblock Performance of hybrid quantum-classical variational heuristics for combinatorial optimization.
\newblock {\em Physical Review E}, 99(1):013304.

\bibitem[Nannicini, 2022]{SimplexQuantum}
Nannicini, G. (2022).
\newblock Fast quantum subroutines for the simplex method.
\newblock {\em Operations Research}.

\bibitem[Pinedo, 2012]{Pinedo}
Pinedo, M.~L. (2012).
\newblock {\em Scheduling}, volume~29.
\newblock Springer.

\bibitem[Ploton and T’kindt, 2022]{Ploton}
Ploton, O. and T’kindt, V. (2022).
\newblock Exponential-time algorithms for parallel machine scheduling problems.
\newblock {\em Journal of Combinatorial Optimization}, 44(5):3405--3418.

\bibitem[Ploton and T’kindt, 2023]{Ploton_flowshop}
Ploton, O. and T’kindt, V. (2023).
\newblock Moderate worst-case complexity bounds for the permutation flowshop scheduling problem using inclusion--exclusion.
\newblock {\em Journal of Scheduling}, 26(2):137--145.

\bibitem[Ruan et~al., 2020]{TSP}
Ruan, Y., Marsh, S., Xue, X., Liu, Z., Wang, J., et~al. (2020).
\newblock The quantum approximate algorithm for solving traveling salesman problem.
\newblock {\em CMC}, 63(3):1237--1247.

\bibitem[Shang et~al., 2018]{Tkindt_3FlowShop}
Shang, L., Lent{\'e}, C., Liedloff, M., and T’Kindt, V. (2018).
\newblock Exact exponential algorithms for 3-machine flowshop scheduling problems.
\newblock {\em Journal of Scheduling}, 21:227--233.

\bibitem[Shimizu and Mori, 2022]{GraphColoring_ProgDyn}
Shimizu, K. and Mori, R. (2022).
\newblock Exponential-time quantum algorithms for graph coloring problems.
\newblock {\em Algorithmica}, pages 1--19.

\bibitem[Sutter et~al., 2020]{ConvexDPQuantum}
Sutter, D., Nannicini, G., Sutter, T., and Woerner, S. (2020).
\newblock Quantum speedups for convex dynamic programming.
\newblock {\em arXiv preprint arXiv:2011.11654}.

\bibitem[Tabi et~al., 2020]{GraphColoring}
Tabi, Z., El-Safty, K.~H., Kallus, Z., H{\'a}ga, P., Kozsik, T., Glos, A., and Zimbor{\'a}s, Z. (2020).
\newblock Quantum optimization for the graph coloring problem with space-efficient embedding.
\newblock In {\em 2020 IEEE (QCE)}, pages 56--62. IEEE.

\bibitem[T’kindt et~al., 2022]{TKindt}
T’kindt, V., Della~Croce, F., and Liedloff, M. (2022).
\newblock Moderate exponential-time algorithms for scheduling problems.
\newblock {\em 4OR}, pages 1--34.

\bibitem[Vazirani, 2001]{VaziraniApprox}
Vazirani, V.~V. (2001).
\newblock {\em Approximation algorithms}, volume~1.
\newblock Springer.

\bibitem[Woeginger, 2003]{Woeginger03}
Woeginger, G.~J. (2003).
\newblock Exact algorithms for np-hard problems: A survey.
\newblock In {\em Combinatorial Optimization—Eureka, You Shrink! Papers}, pages 185--207. Springer.

\end{thebibliography}

\appendix

\section{Useful upper bounds}
\label{appendix:Bounds}

We define the binary entropy of $\epsilon\in ]0,1[$ by $H(\epsilon) = -(\epsilon\log_2(\epsilon) + (1-\epsilon)\log_2(1-\epsilon))\,.$
We remind some useful upper bounds of binomial coefficients~\cite{Ambainis}:
$$
\tiny
\binom{n}{k} \leq 2^{H\left(\frac{k}{n}\right)\cdot n},\;\,\forall k\in\llbracket 1,n\rrbracket
\qquad\mbox{and}\qquad
\sum_{i=1}^{k}\binom{n}{i} \leq 2^{H\left(\frac{k}{n}\right)\cdot n},\,\forall k\in\left\llbracket 1,\frac{n}{2}\right\rrbracket.
$$
Thus, it leads to the following upper bounds to compute the complexities of interest: 
\begin{equation*}
\tiny
    \sum_{i=k}^{n/4}\binom{n}{k} \leq  2^{0.811n},
    \sum_{k=1}^{0.945\cdot n/4}\binom{n}{k} \leq 2^{0.789n},
    \sqrt{\binom{n}{n/2}\binom{n/2}{n/4}} \leq 2^{0.75n},\sqrt{\binom{n}{n/2}\binom{n/2}{n/4}\binom{n/4}{0.945\cdot n/4}} \leq 2^{0.789n} 
\end{equation*}
\end{document}